\documentclass[11pt]{article}

\usepackage[T1]{fontenc}
\usepackage{authblk}

\usepackage{hyperref}
\usepackage{verbatim}
\usepackage{amsmath}
\usepackage{graphicx}

\usepackage{makecell}

\usepackage{subcaption}
\usepackage{cleveref}

\usepackage{array}

\usepackage{relsize}
\usepackage{colortbl}

\usepackage{setspace}

\usepackage{url}
\usepackage{tablefootnote}

\usepackage{color}

\usepackage{color, colortbl}
\definecolor{Gray}{gray}{0.9}
\definecolor{LightCyan}{rgb}{0.88,1,1}
\definecolor{Pink}{rgb}{1,0.75,0.70}
\definecolor{Gray}{rgb}{0.84,0.84,0.84}

\pdfoutput=1

\begin{document}


\title{Gender Disparities in Science? Dropout, Productivity, Collaborations and Success of Male and Female Computer Scientists}
\date{}

\author[1]{Mohsen Jadidi}
\author[1]{Fariba Karimi}
\author[1]{Haiko Lietz}
\author[1]{Claudia Wagner}

\affil[1]{GESIS - Leibniz Institute for the Social Sciences, Cologne, Germany \\ Email:\{firstname.lastname\}@gesis.org }




\maketitle

\begin{abstract}
Scientific collaborations shape ideas as well as innovations and are both the substrate for, and the outcome of, academic careers.
Recent studies show that gender inequality is still present in many scientific practices ranging from hiring to peer-review processes and grant applications.
In this work, we investigate gender-specific differences in collaboration patterns of more than one million computer scientists over the course of 47 years. We explore how these patterns change over years and career ages and how they impact scientific success.
Our results highlight that successful male and female scientists reveal the same collaboration patterns: compared to scientists in the same career age, they tend to collaborate with more colleagues than other scientists, seek innovations as brokers and establish longer-lasting and more repetitive collaborations.
However, women are on average less likely to adapt the collaboration patterns that are related with success, more likely to embed into ego networks devoid of structural holes, and they exhibit stronger gender homophily as well as a consistently higher dropout rate than men in all career ages.

\end{abstract}


Collaboration in social networks is how scientists collectively negotiate the direction of research in a field or discipline \cite{white_identity_2008}.

In the course of a collaboration, new ideas shape and eventually result in new discoveries and scientific publications \cite{moody2004structure}.
As a result, collaborations impact researchers' scientific careers and their academic success \cite{petersen2014reputation,petersen_tie,Sarigoel2014,servia2015evolution}.
For example, the centrality of a scientist in a collaboration network is associated with her success \cite{Sarigoel2014,servia2015evolution} and co-authorship strength is related to high productivity and citations \cite{petersen_tie,petersen2014reputation}.

Gender inequality is still rife in science.
Previous work on scientific collaboration suggest that men and women tend to exhibit different collaborative behaviours across their scientific career, such as differences in the number of new and repeating collaborations \cite{kyvik1996child,Xiao2016}, position in the network \cite{West2013}, degree of homophilic behaviour \cite{ECIN:ECIN68,gallivan2015co} and tendency to have interdisciplinary collaborations \cite{Leahey01122006,Rhoten200756}.

The \emph{productivity puzzle} refers to the unknown causes of the lower publication rate of women compared to men in various fields \cite{cole_1984}. Many studies have provided explanations of possible underlying causes of this productivity gap in science \cite{West2013,bentley_2003,wenneras2001nepotism,Duch2012,cole_1984,prozesky2008career,Stack2004}. For example, Duch et al. found that women publish significantly fewer papers in fields where research is expensive \cite{Duch2012}. Following this argument, differences in research funding could be one factor behind the productivity puzzle. Other studies point towards family responsibilities \cite{prozesky2008career,Stack2004}, discrimination in the peer-review process \cite{wenneras2001nepotism,knobloch2013matilda}, the employment position (i.e. being a professor or a post-doc) \cite{bentley_2003} and international collaboration \cite{Prpi2002} as factors that may explain the lower productivity of women.

Gender differences have also been observed in hiring \cite{Moss2012}, grant applications \cite{Ley2008,vanDerLee2015}, peer reviews \cite{doi:10.1027/2151-2604/a000103,Kaatz2014371}, earnings \cite{Holden2001,West2006}, tenure \cite{Spelke05}, satisfaction \cite{Holden2001}, patenting \cite{Ding2006}, scientific success \cite{Cassidy2013}, collaborations \cite{Cassidy2013,Xiao2016} and division of labour in scientific collaborations \cite{Macaluso2016}. 
For example, a report from 2006 showed that only one quarter of full professors are female and that they earn 80\% of their male colleagues' wages on average \cite{West2006}. More recent research showed that women are more likely to take executive roles in collaborations \cite{Macaluso2016}, their collaborations are more domestically oriented and papers with women as lead author (i.e. solo author, first author or last author) receive fewer citations \cite{Cassidy2013}.
A recent study that investigated collaboration patterns of female and male researchers in science, technology, engineering and mathematical (STEM) disciplines found that female scientists have significantly fewer distinct co-authors over their careers and a lower probability of repeating previous collaborations than males \cite{Xiao2016}.

In this work we extend these lines of research by investigating collaboration patterns of male and female scientists. Unlike previous work, we analyze the temporal evolution of collaborations in one entire field, computer science, and compare the structural position and the success of men and women over their career ages. We use the number of citations and the $h$-index to operationalize the success of scientists and explore to what extent the position of men and women in their networks explains their success and if there are gender-specific differences. A solution of the productivity puzzle is sought and homophily is studied.

Our main results are that
(1) the dropout rate of women is consistently higher than of men, especially at the beginning of an academic career;
(2) the productivity puzzle can be solved and explained by the higher number of senior male scientists;
(3) there is a sizeable and constant division of labour in the sense that the ego networks of female researchers are much more closed and contain fewer brokerage opportunities;
(4) closure and brokerage are co-determinants of scientific success, there are no gender-specific differences in how collaborative behaviour impacts scientific success, but men are more likely to adapt the collaborative practices that are related to success; and
(5) gender homophily has been increasing over the past few years.

\section{Data}

\begin{table}[t]
\centering
\caption{The proportion of correct guesses for various gender detection methods for scientists across different countries. For most countries the mixed approaches that combine name- and image-based gender detection perform best. \label{tab:accuracy_country}}

\resizebox{\columnwidth}{!}{%
\begin{tabular}{|l|c|c|c|c|c|c|c|c|}
\hline
               & \# instances & SSA  & IPUMS & Sexmachine              & Genderize & Face++                 & Mixed1               & Mixed2                  \\ \hline
United States  & 419         & 0.82 & 0.76  & 0.84   & 0.83  & 0.91    & \cellcolor{Gray} 0.91 & 0.90                    \\ \hline
China          & 113         & 0.20 & 0.11  & \cellcolor{Gray} 0.67 & 0.28      & 0.65                      & 0.50                    & 0.56                    \\ \hline
United Kingdom & 96          & 0.94 & 0.92  & 0.92                    & 0.94      & 0.81                    & \cellcolor{Gray} 0.98 & 0.94                    \\ \hline
Germany        & 82          & 0.87 & 0.88  & \cellcolor{Gray} 0.96 & 0.94      & 0.87                      & \cellcolor{Gray} 0.96 & 0.93                    \\ \hline
Italy          & 75          & 0.93 & 0.92  & 0.94                    & 0.98      & 0.79                    & 0.99                    & \cellcolor{Gray} 1    \\ \hline
Canada         & 60          & 0.87 & 0.77  & 0.86                    & 0.91      & 0.90                    & \cellcolor{Gray} 0.96 & 0.93                    \\ \hline
France         & 58          & 0.93 & 0.92  & 0.80                    & 0.96      & 0.81                    & 0.97                    & \cellcolor{Gray} 1    \\ \hline
Japan          & 56          & 0.79 & 0.70  &\cellcolor{Gray} 1                    & 0.90      & 0.62                    & 0.91                    & \cellcolor{Gray} 0.94 \\ \hline
Brazil         & 44          & 0.29 & 0.29  & 0.15                    & 0.44      & 0.81                    & 0.90                    & \cellcolor{Gray} 0.93 \\ \hline
Spain          & 39          & 0.96 & 0.92  & 0.92                    &\cellcolor{Gray} 1      & 0.92                    & \cellcolor{Gray} 1    & \cellcolor{Gray} 1    \\ \hline
Australia      & 31          & 0.89 & 0.89  & 0.90                    & 0.86      & 0.86                    & \cellcolor{Gray} 0.94 & 0.93                    \\ \hline
India          & 29          & 0.67 & 0.17  & 0.71                    & 0.78      & 0.83                    & 0.83                    & \cellcolor{Gray} 0.93 \\ \hline
South Korea    & 27          & 0.04 & 0.00  & 0.58                    & 0.11      & \cellcolor{Gray} 0.74   & 0.37                    & 0.66                    \\ \hline
Switzerland    & 25          & 0.78 & 0.70  & 0.56                    & 0.83      & 0.88                    & 0.90                    & \cellcolor{Gray} 0.92 \\ \hline
Turkey         & 21          & 0.43 & 0.14  & 0.79                    & 0.81      & 0.86                    & \cellcolor{Gray} 1    & \cellcolor{Gray} 1    \\ \hline
\end{tabular}
}
\end{table}

To construct a time-evolving collaboration network we use the DBLP Computer Science Bibliography \cite{DBLP:journals/pvldb/Ley09}, a comprehensive collection of computer science publications from major and minor journals and conferences. While DBLP offers name disambiguation \cite{DBLP:series/lnsn/Reitz013,DBLP:journals/pvldb/Ley09,DBLP:conf/ercimdl/ReutherWLWK06}, it does not provide information about citations. Therefore, we use publication titles to combine the DBLP dataset with the Aminer dataset \cite{Tang:08KDD} that contains all citation relations among papers in DBLP. 

To infer the gender of authors we utilized a method from a previous study that combines the result of name-based (Genderize.io\footnote{\url{https://genderize.io/}}) and image-based (Face++\footnote{\url{https://www.faceplusplus.com/}}) gender detection services \cite{Karimi:2016}. Compared to other name-based methods, our approach achieves a high accuracy (above 90\%) for most countries (see \emph{Mixed1} in Table \ref{tab:accuracy_country}). For evaluation we used ground-truth data from a previous study that was manually compiled by looking at the CVs, pictures and institutional websites of a random sample of scientists (693 men and 723 women) \cite{Cassidy2013}.

Our combined approach does a better job in inferring non-western names but also performs poorly for Asian (Chinese or Korean) names (see Table \ref{tab:accuracy_country}). 
Therefore, we first detect which names are Asian and label their gender as ``unknown''.
The authors for which we cannot detect the gender are excluded from our gender-specific analyses, but included in all structural analyses (e.g., as alters).

To detect Chinese names we compile a list of 202,045 unique names from the China Biographical Database Project (CBDB)\footnote{\url{http://projects.iq.harvard.edu/cbdb/home}}. For compiling a list of Korean names we use Wikipedia as our data source. To do this, we extract the page titles of all the backlinks to the Wikipedia page ``Korean names''\footnote{\url{https://en.wikipedia.org/wiki/Korean_name}}. The page titles include the names of prominent Korean figures (e.g., singers) with a Wikipedia page that describe the origin of the name of that person (e.g., Wikipedia page of a Korean singer and actor\footnote{\url{https://en.wikipedia.org/wiki/Ahn_Jae-wook}}). Using this method we compile a list of 6,451 unique Korean names. 
A manual evaluation of our Asian name detector shows that 88 out of 100 randomly selected scientists were correctly classified; we found 20 true positives (Asians classified as Asians), 68  true negatives (Non-Asians classified as Non-Asians), 2 false negatives (Asians classified as Non-Asians) and 10 false positives (Non-Asians classified as Asians). The relatively high number of false positives can be explained by the fact that some famous Asians choose western names.
In addition, we exclude authors with only first initials since we cannot infer the gender of authors without knowing their first name. This may exclude female authors disproportionately, particularly in early decades when women may have been more likely than men to publish with initials to avoid potential discrimination. But we expect that this difference is small since our data collection only goes back to 1970.

\begin{figure}[t]
\centering
\begin{minipage}{0.47\linewidth}
\centering
  \includegraphics[width=\textwidth]{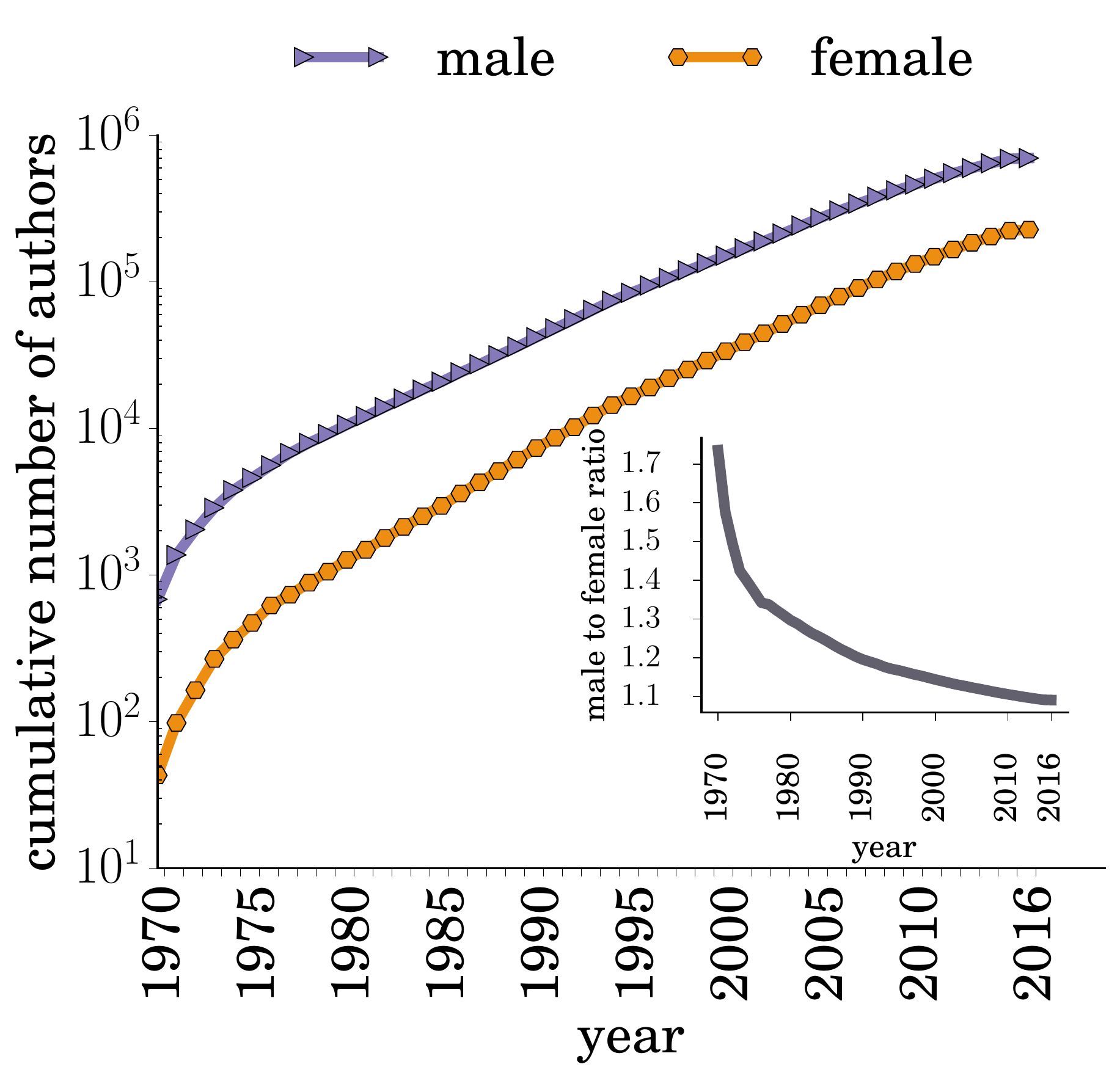} 
\end{minipage}
\begin{minipage}{0.47\linewidth}
\centering
\includegraphics[width=\textwidth]{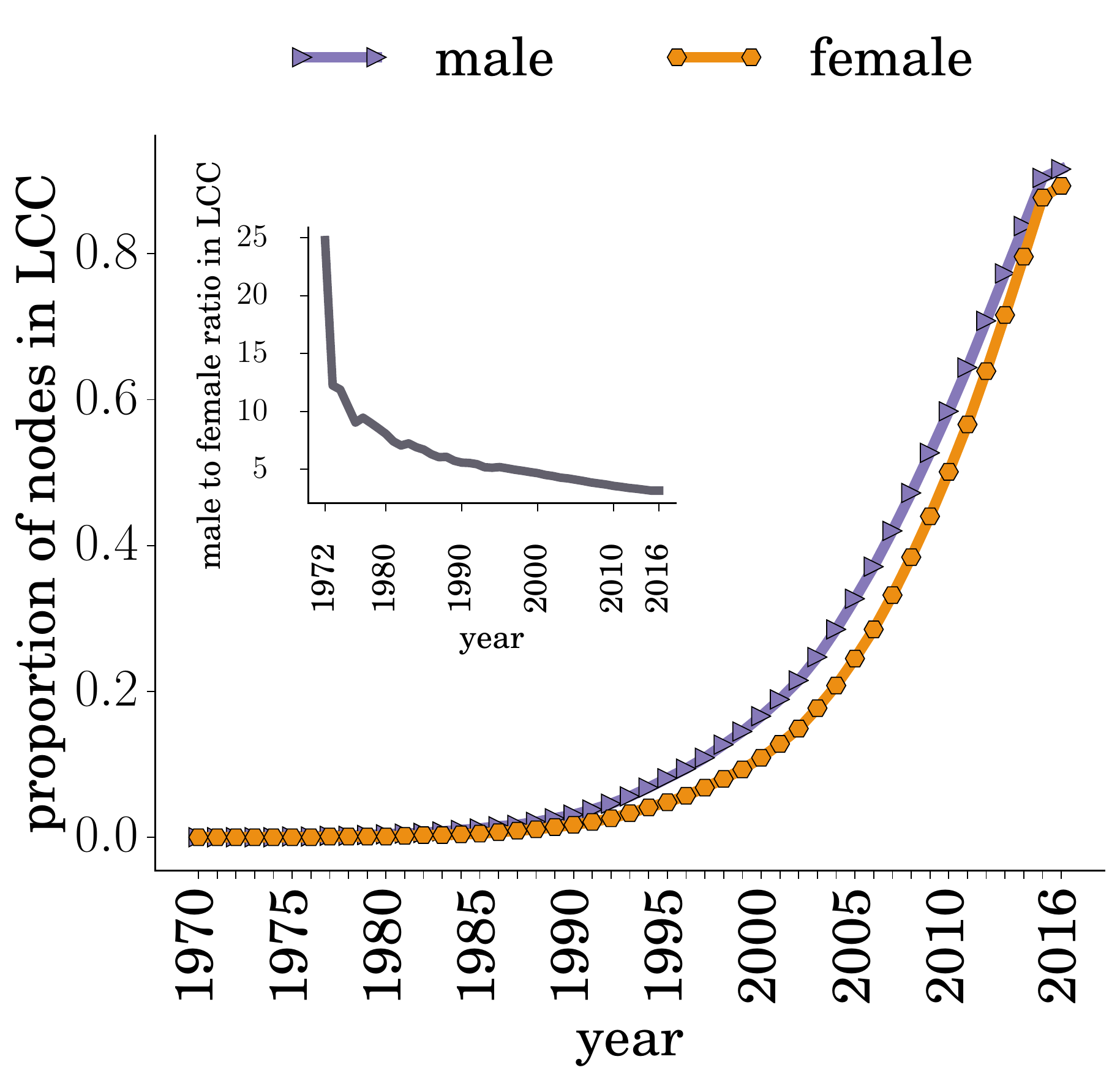}
\end{minipage}
\caption{{\textbf{Left: Presence of men and women in the community.} The main figure shows the cumulative number of men and women. The inset shows the corresponsing ratio of men to women. Women are always underrepresented in the community, but the gap is closing. \textbf{Right: Growth of Largest Connected Component (LCC).} The main figure shows the proportion of men and women that belong to the LCC of the cumulative network.
There is always a higher proportion of men that belong to the LCC. The inset shows that the men/women ratio in the LCC is also decreasing over time. There are no women in the LCC in 1970 and 1971.}
 \label{fig:net_stats}}
\end{figure}

Our dataset consists of 1,634,682 scientists, 3,085,544 publications and 7,849,398 citations that have been created in the time span of 47 years, between 1970 and 2016. Among all publication, 717,471 papers (23\%) receive at least one citation from other papers inside the DBLP corpus.

For all authors with known and unknown gender we build a collaboration network where each node represents an author and each edge a co-authorship relation.
The complete graph consists of 1,634,682 nodes and 7,304,250 edges. 699,370 (43\%) authors were identified as men, 227,473 (14\%) as women, and for 707,839 (43\%) authors gender in unknown.
Each edge is labeled by one or multiple date(s) that correspond to the publication year(s) of papers. We later use this information to study the network's evolution over time.

We take a career approach, i.e. we study researchers at multiple steps in their career.

We infer the \emph{career ages} of scientists by comparing their first and last publication record inside the DBLP corpus. For example, a scientist who has only published papers in 1995, 2000 and 2005 has a career length of 11 years. In 1995 her career age is 1, in 2000 it is 6 and in 2005 it is 11. 

\textbf{Descriptive statistics.} Figure~\ref{fig:net_stats} (left) shows that the computer science community has been growing rapidly in recent years and is becoming more gender-balanced. The inset suggests the gender gap is closing over time where the men/women ratio decreases from 1.7 in 1970 to 1.1 up until 2015.

Figure~\ref{fig:net_stats} (right) depicts the proportion of men and women that are part of the Largest Connected Component (LCC). For example, in 2000, about 20\% of men and  10\% of women were part of the LCC. Until 2015, the proportion of scientists in the LCC had increased to about 85\% and 80\%, respectively.

This increase resembles an increase of network connectivity which is partly due to the cumulative construction of the graph and partly due to endogenous densification \cite{yang2012structure}. 
The recency bias in the coverage of DBLP \cite{Way2017} may add to the observation of the growing LCC. However, this bias should equally effect publications of men and women, and relative differences between men and women should, therefore, still be meaningful.

Essentially, the inset reveals that the proportion of men in the LCC has always been higher than those of women. However, the gap is closing over time.

\section{Results}

To investigate the evolution of gender disparities in the computer science community between 1970 and 2015, we compare  (1) \emph{dropouts} (number of male and female scientists that stop publishing), (2) \emph{productivity} (number of publications per author), (3) \emph{collaboration patterns} and (4) \emph{scientific success} (number of citations and h-index) of male and female scientists.

\subsection{Dropout}
\begin{figure}[ht!]
\centering
  \includegraphics[width=0.6\linewidth]{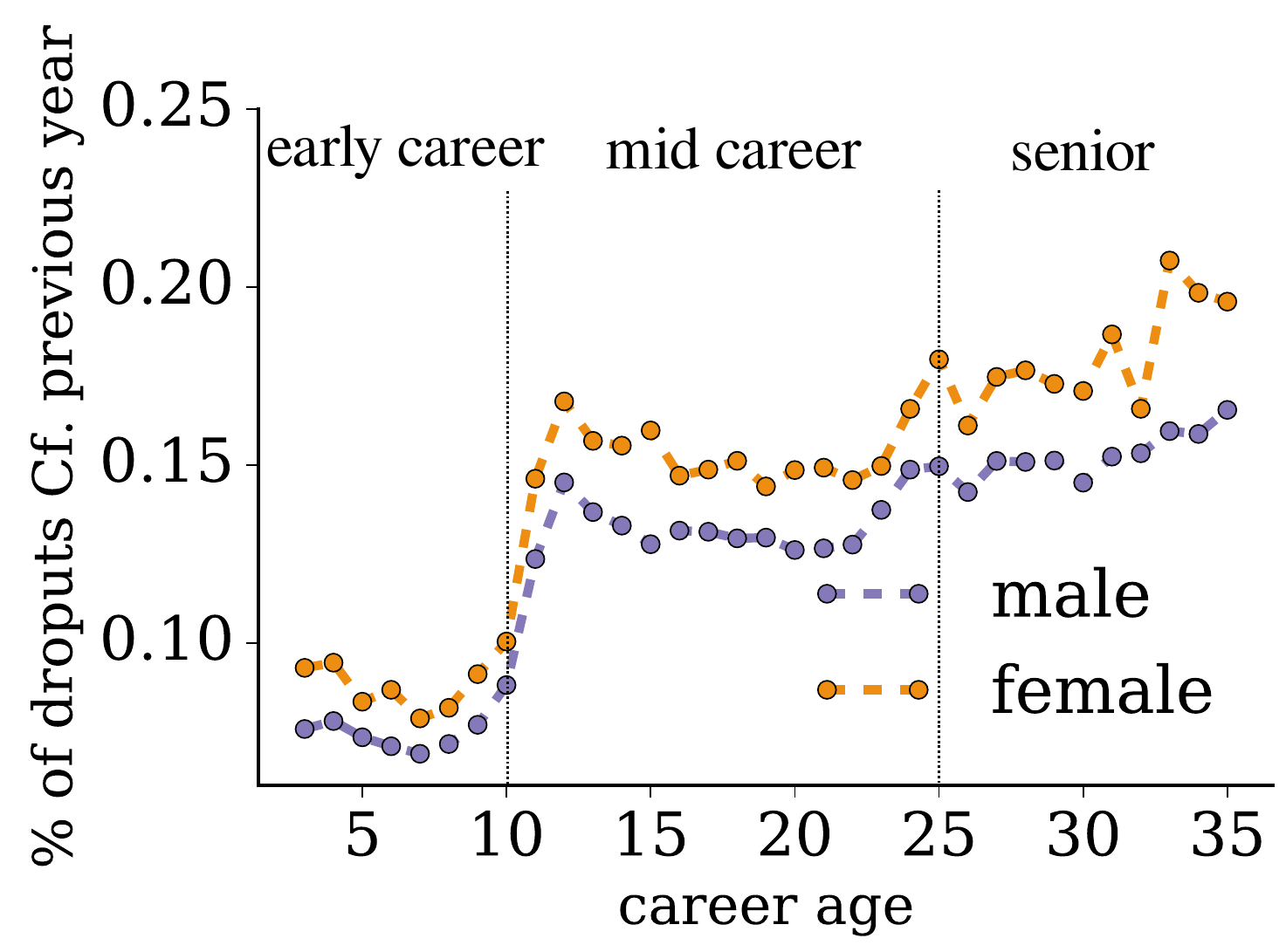}
\caption{\textbf{Dropout rate:}
Proportion of men and women at different career ages that permanently stop publishing. Most scientists (40\% of men and 47\% of women) drop out one year after their first publication (not shown). Of those that continue, 8\% of men and 9\% of women drop out after their second year (from here on shown).
After the drastic dropout at the very beginning, the rate shows three phases. The first corresponds to early-career researchers (career age 2-10) for which we observe a dropout rate between 7\% and 10\% every year.
In career ages 11 and 12, the rate jumps to 15\% for men and 17\% for women. In the second phase related to mid-career researchers (career age 11-25), the dropout rate fluctuates between 13\% and 18\%.
The third phase corresponds to senior researchers with (career age above 25). They drop out at a rate of 14\% to 21\% (for career age above 35 fluctuations increase). Women consistently have higher rates (2 percentage points) across all career ages.
\label{fig:dropout}}
\end{figure}

\emph{Leaky pipelines} are frequently claimed to cause gender disparities in science. This metaphor implies that women drop out of academia at a higher rate as they advance in their career \cite{wickware1997along,pell1996fixing}. To compare the dropout rates of male and female scientists we first infer their career age based on their publications.
We assume that a scientist who has not published any paper in 10 or more years has left academia, since staying in academia requires publishing. 
Scientists who died will also be counted as dropouts, but we do not expect that the proportion of men and women who die in the same career age is significantly different.
Since our dropout definition requires to observe at least 10 years after each publication, we limit our dataset to scientists who published at least one publication before 2006.
That means people who started their scientific career after 2006 are not included in our analysis. This leaves 326,329 men and 84,859 women for the dropout analysis.

Figure \ref{fig:dropout} shows the percentage of men and women who permanently dropped out of the academic pipeline at different stages in their academic career. The main message is that scientists tend to stay in the field if they manage to survive the first year in which they publish. 40\% of the male and 47\% of the female authors do not enter a second year (read caption for further details). For those who do survive,
32\% of the men and 31\% of the women stay for up to 10 years and become \emph{early-career researchers},
25\% of the men and 20\% of the women stay for up to 25 years and become \emph{mid-career researchers},
and only 3\% of the men and 2\% of the women become \emph{senior researchers} and stay 26 and more years in the field.
This gender difference of careers entails a comparability issue we need to address in the remainder of the paper.

\begin{figure*}[t!]
\centering
\begin{minipage}{.31\linewidth}
\centering
  \includegraphics[width=\textwidth]{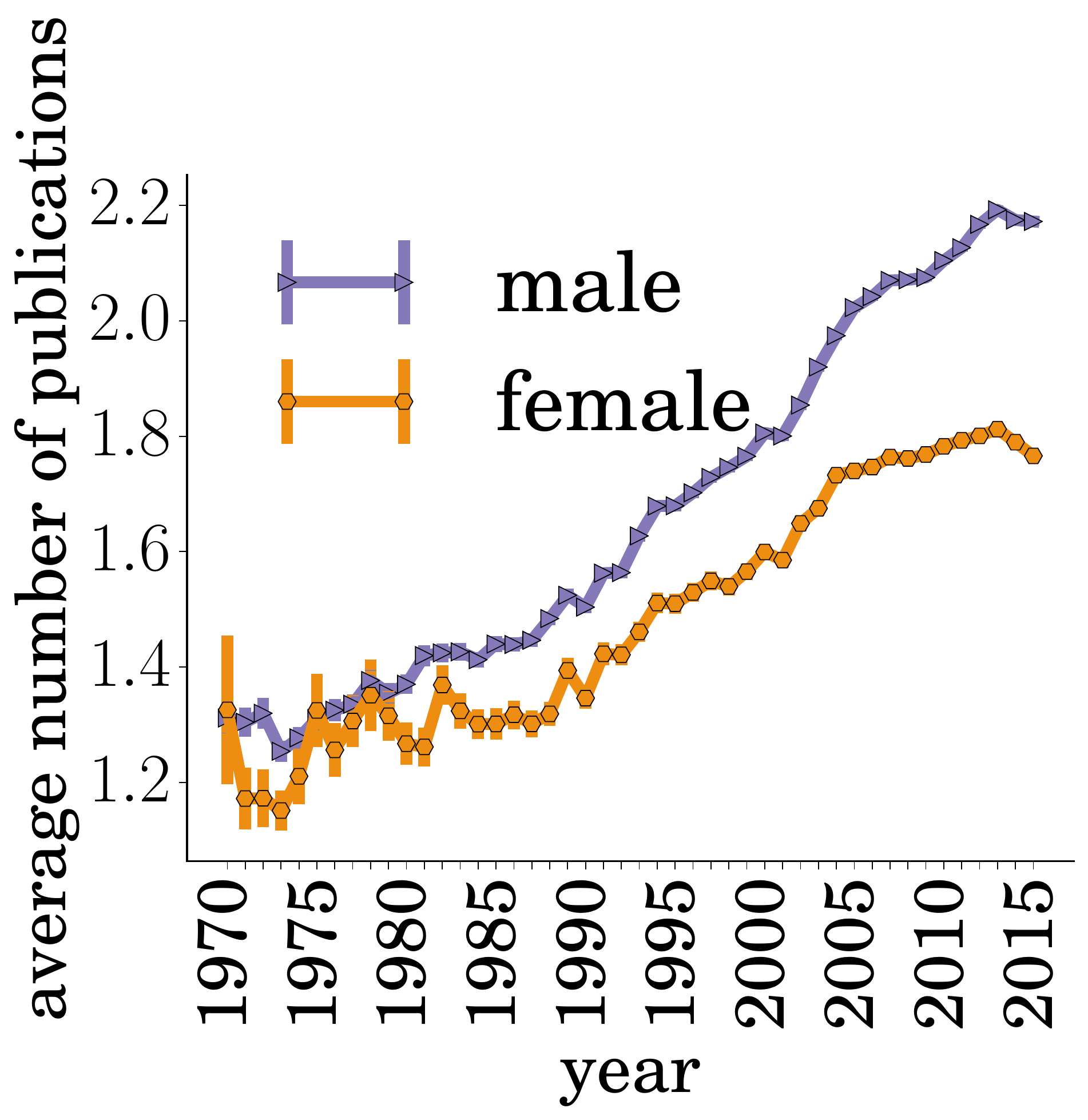}
\end{minipage}
\begin{minipage}{.31\linewidth}
\centering
  \includegraphics[width=\textwidth]{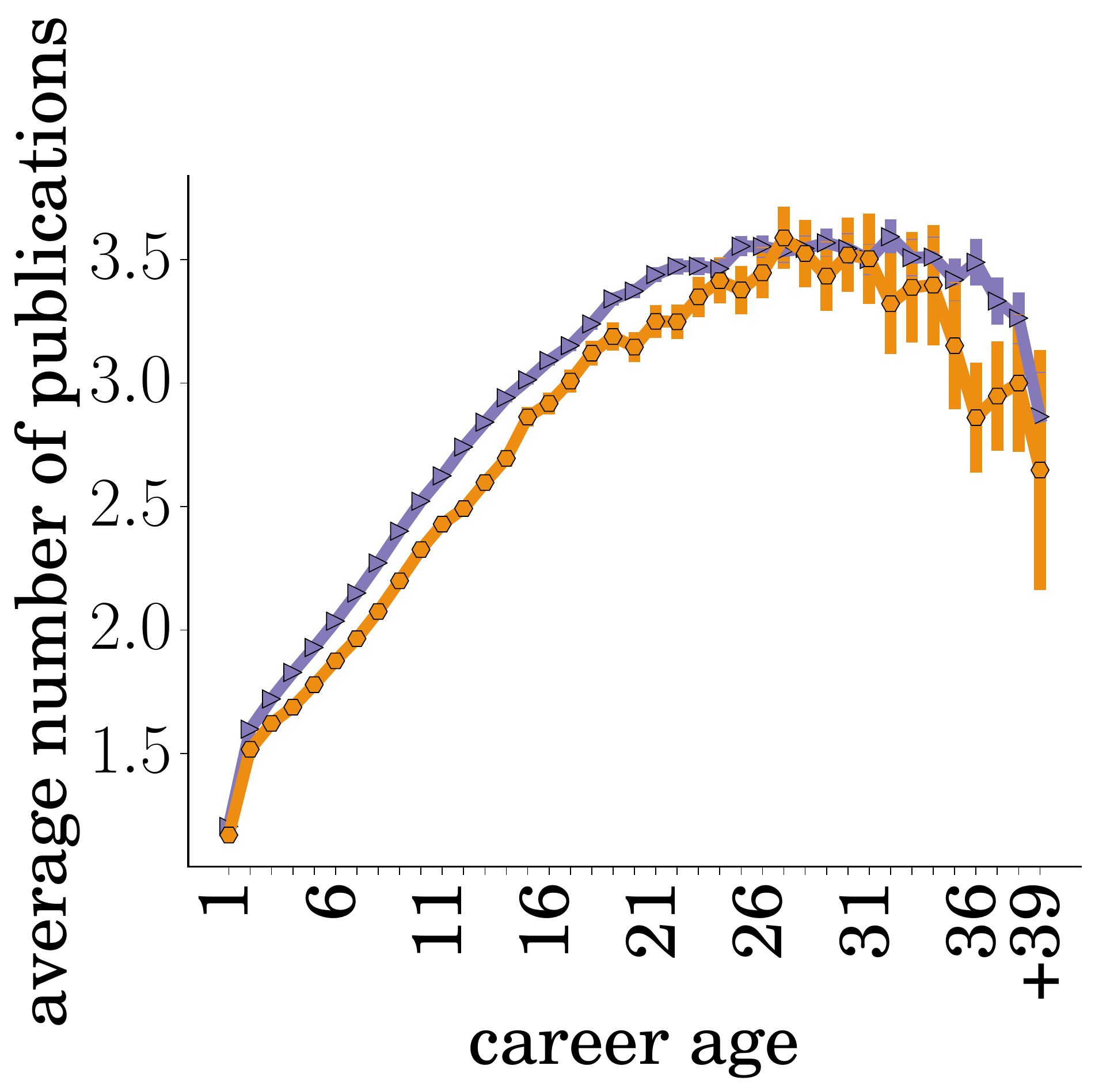}
\end{minipage}
\begin{minipage}{0.31\linewidth}
\centering
  \includegraphics[width=\textwidth]{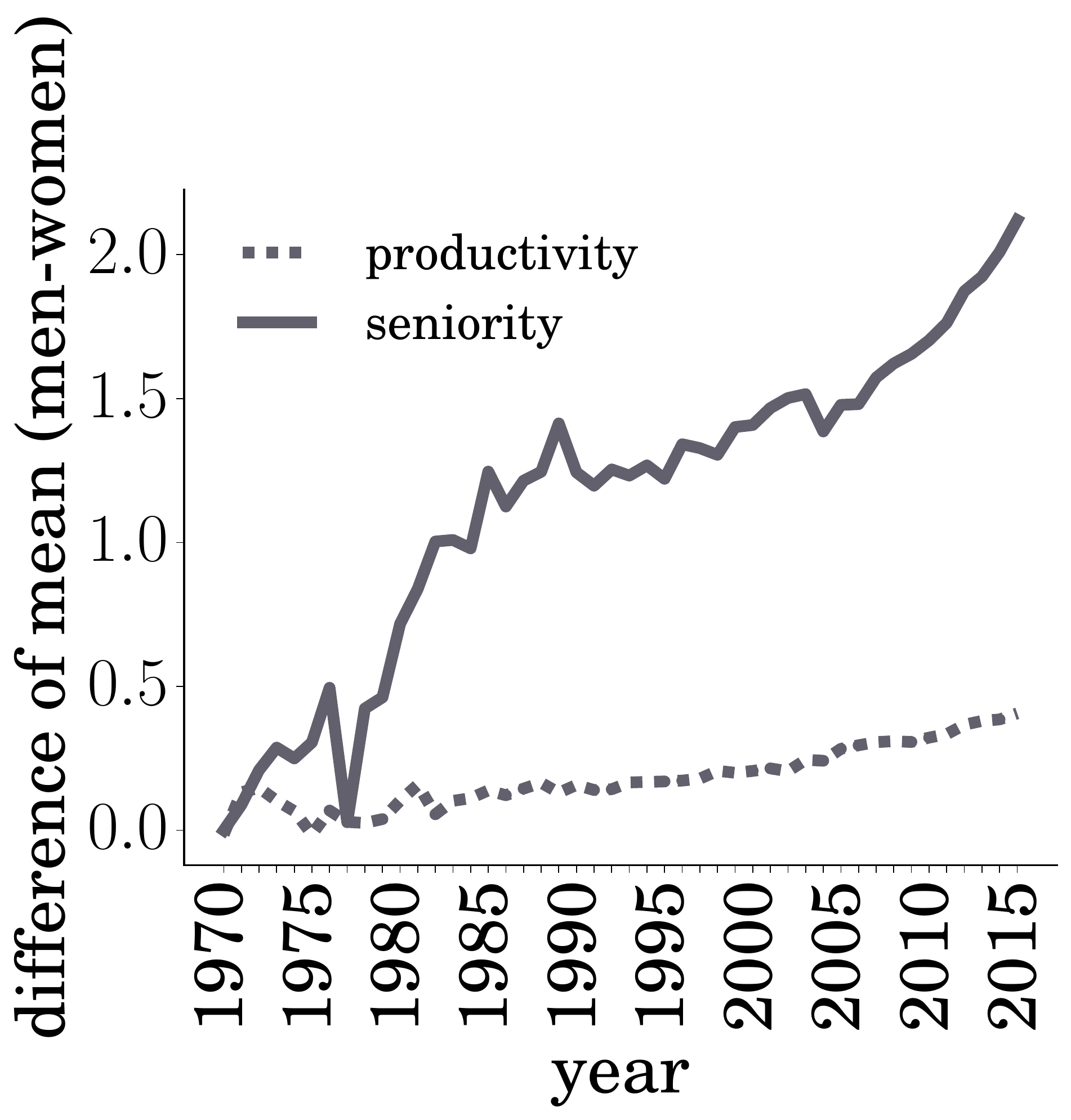}
\end{minipage}
\caption{
\textbf{Left: Productivity gap (calendar years).} Average productivity (number of publications) of men and women over calendar years. Although productivity increases for both sexes, men tend to be slightly more productive than women. In this analysis we neglect the year 2016 as it might be affected by censoring bias and missing publications.
\textbf{Middle: Productivity gap (career ages).} Average productivity of men and women over career ages.
Three phases can roughly be detected: (1) career age 1-20: increase of productivity; (2) career age 21-30: stable productivity, 3) career age 31 and on: decreases of productivity. The average productivity of men and women at the same stage of the career is very similar.
\textbf{Right: Productivity gap vs. seniority gap.}  Differences between the mean productivity of men and women (productivity gap) and the mean career ages of men and women (seniority gap) in the same calendar year. The Pearson correlation between the two differences is 0.86 with $p=10^{-15}$.
\label{fig:pub_age}}
\end{figure*}

\subsection{Productivity}
Various explanations, from funding to family responsibilities and international collaboration, have been offered to solve the productivity puzzle discussed in the introduction.
Our results show that the average productivity, regardless of gender, has been increasing over time and that gender differences prevail (cf. figure \ref{fig:pub_age}, left). On average, men tend to have higher publication rates than women in all calendar years and the gap is widening after 2005.

We offer a solution to the productivity puzzle.
The productivity gap almost vanishes when the average productivity of men and women in the same career age is compared (cf. figure \ref{fig:pub_age}, middle).
Three phases of productivity become very similar for men and women:
In their first two decades scientists tend to increase their productivity each year. In the following 10 years their average productivity is rather stable and scientists produce about 3.0 to 3.5 publications per year on average. Towards the end of long careers productivity drops again.

This result is in line with previous studies that found a similar pattern of productivity over the chronological age of scientists \cite{phelan,ASI:ASI21486,Vasileiadou20091260,Lehman1954}. However, the literature also reports different productivity trajectories for scientists of different citation impact \cite{Sinatraaaf5239} and for researchers in different disciplines \cite{Bayer1977,kyvik1996child}. Recent research also highlights that while the aggregated pattern of productivity is surprisingly similar for researchers that are placed in institutions of different prestige rank, high diversity can be observed in the production trajectories of individual scientists \cite{Way2017}.

Comparing scientists only for similar career ages amounts to controlling for seniority. Figure \ref{fig:pub_age} (right) shows that the \emph{productivity gap}, measured as the difference between the mean productivity of men and women in the same year, is paralleled by a \emph{seniority gap}, measured as the difference between the mean career age of men and women in the same year. They not only increase over time but are strongly and significantly correlated (Pearson correlation coefficient $0.86$, $p = 10^{-15}$).
This suggests the simple explanation that men are more productive on average because they have a larger fraction of senior authors.

\subsection{Collaboration patterns}

Previous studies have either focused on a specific country (e.g., Zeng et al. \cite{Xiao2016} focus on the US) or ignored the time dimension (e.g., West et al. \cite{West2013} ignore the career age of men and women when analyzing the average authorship-position on papers).

Here we investigate \emph{how collaboration patterns and the network positions of male and female researchers change over time in an entire scientific field, computer science}.
\begin{figure*}[t]
\begin{subfigure}{.3\linewidth}
  \centering
  \includegraphics[width=\textwidth]{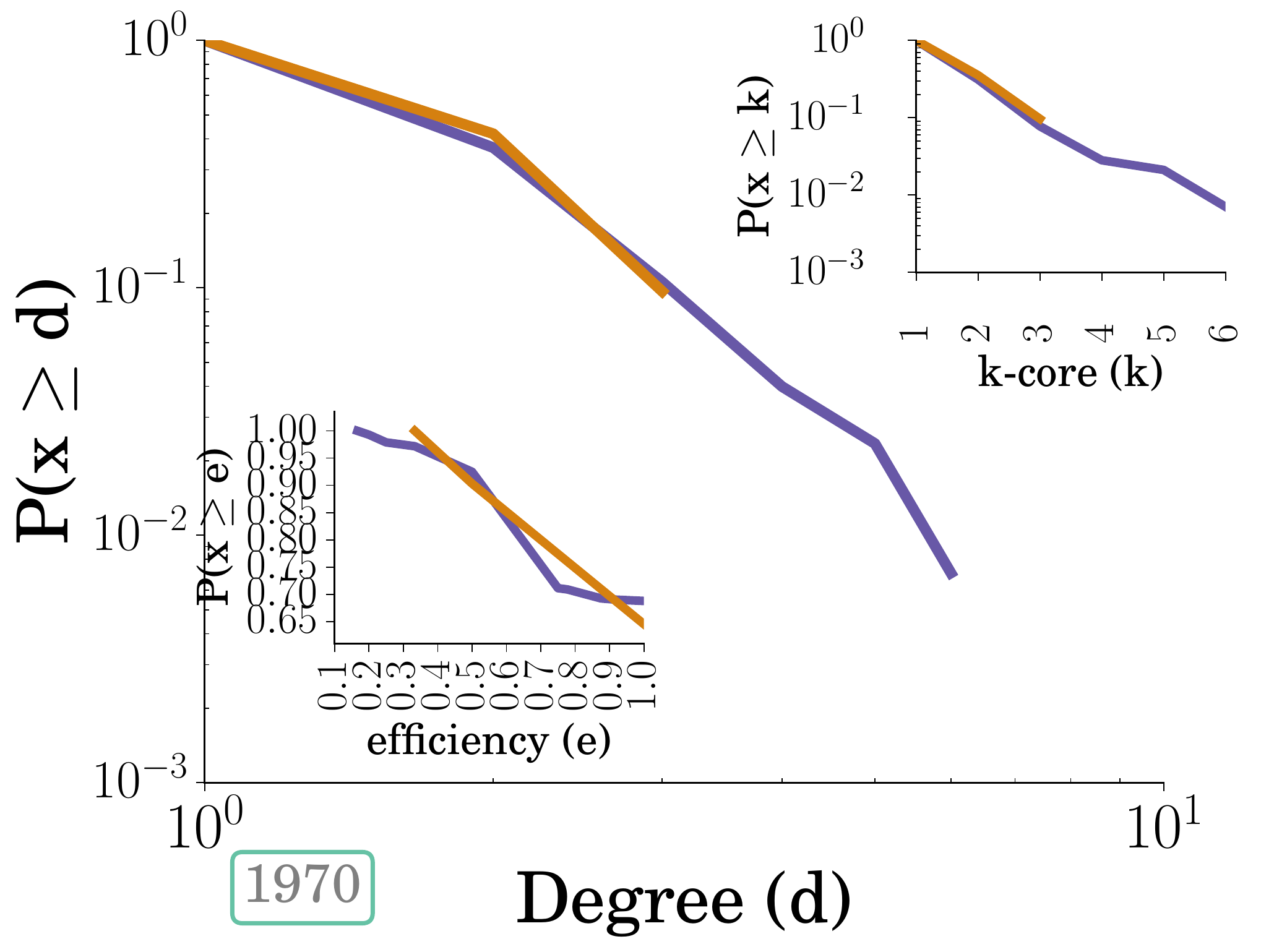}\quad
\end{subfigure}
\begin{subfigure}{.3\linewidth}
  \centering
  \includegraphics[width=\textwidth]{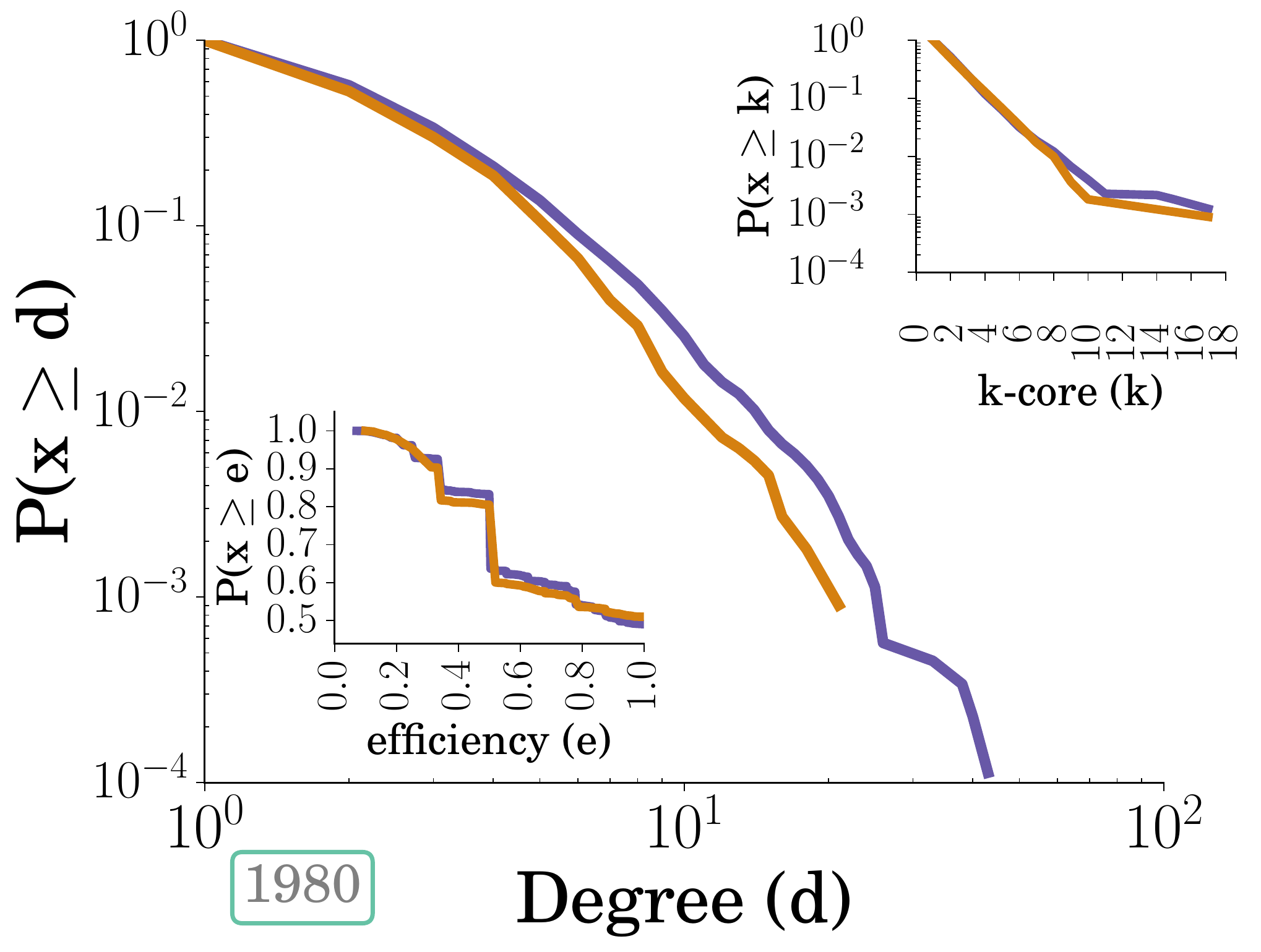}\quad
\end{subfigure}
\begin{subfigure}{.3\linewidth}
  \centering
  \includegraphics[width=\textwidth]{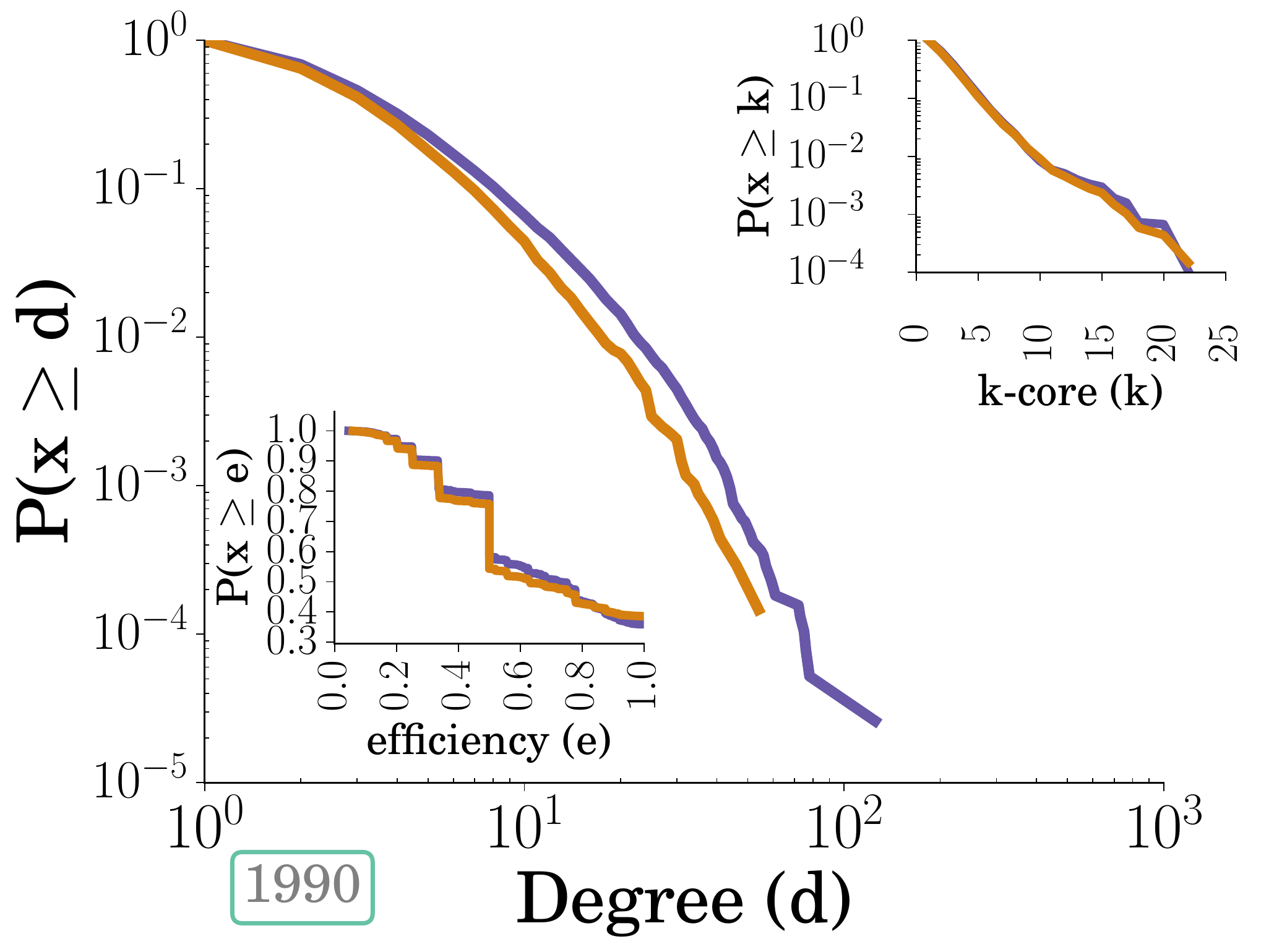}
\end{subfigure}

\medskip
\begin{subfigure}{.3\linewidth}
  \centering
  \includegraphics[width=\textwidth]{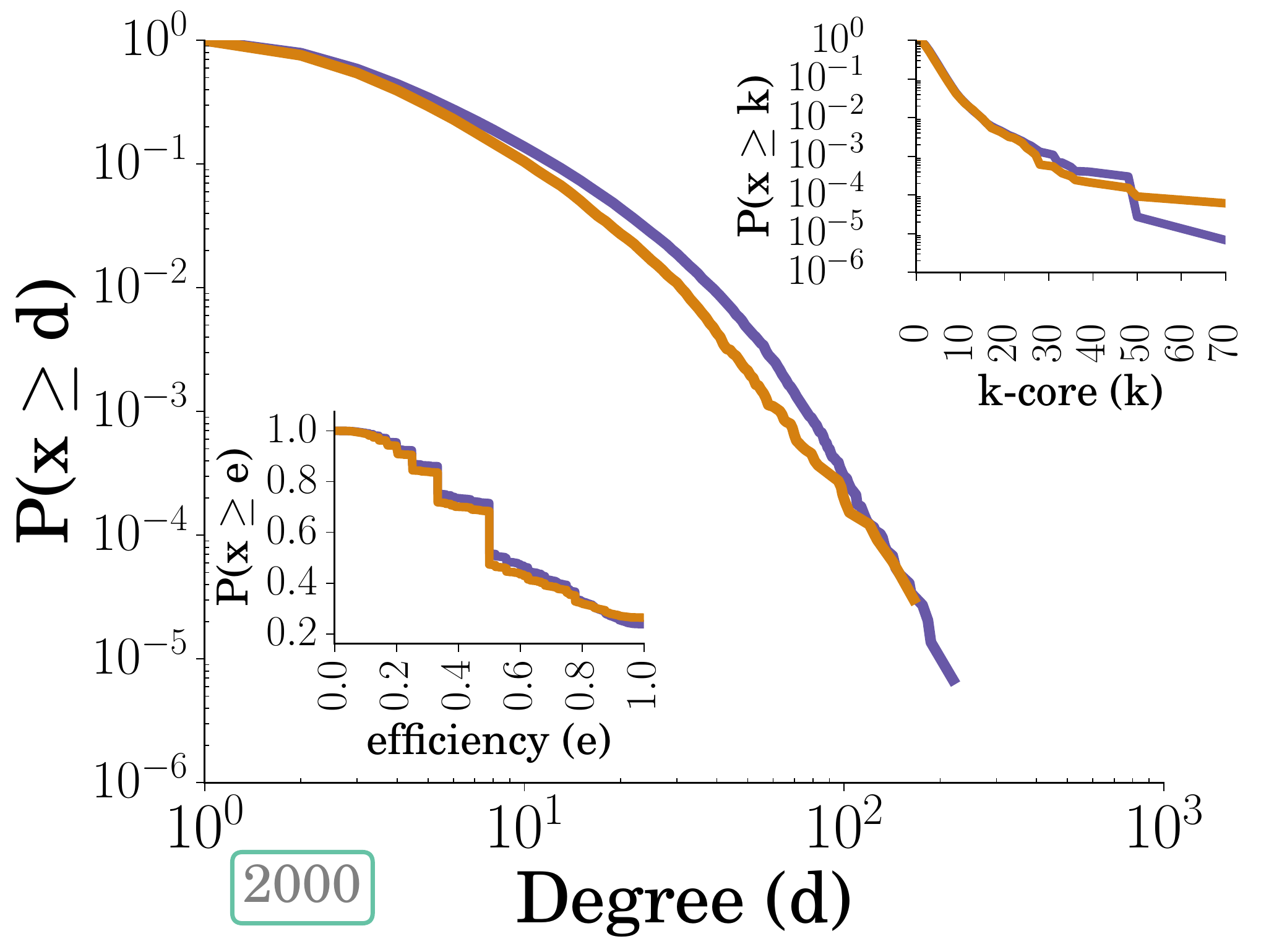}\quad
\end{subfigure}
\begin{subfigure}{.3\linewidth}
  \centering
  \includegraphics[width=\textwidth]{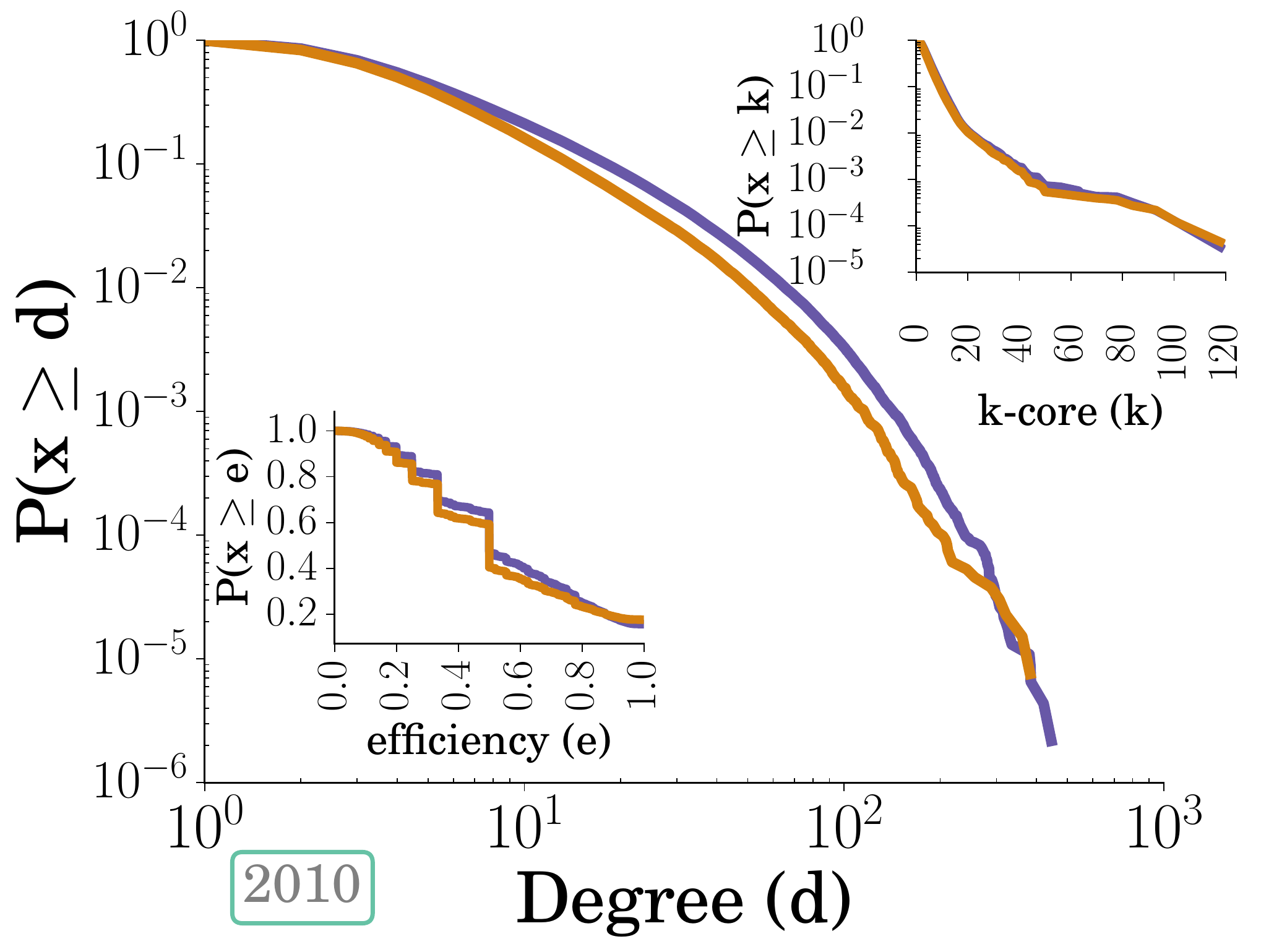}\quad
\end{subfigure}
\begin{subfigure}{.3\linewidth}
  \centering
  \includegraphics[width=\textwidth]{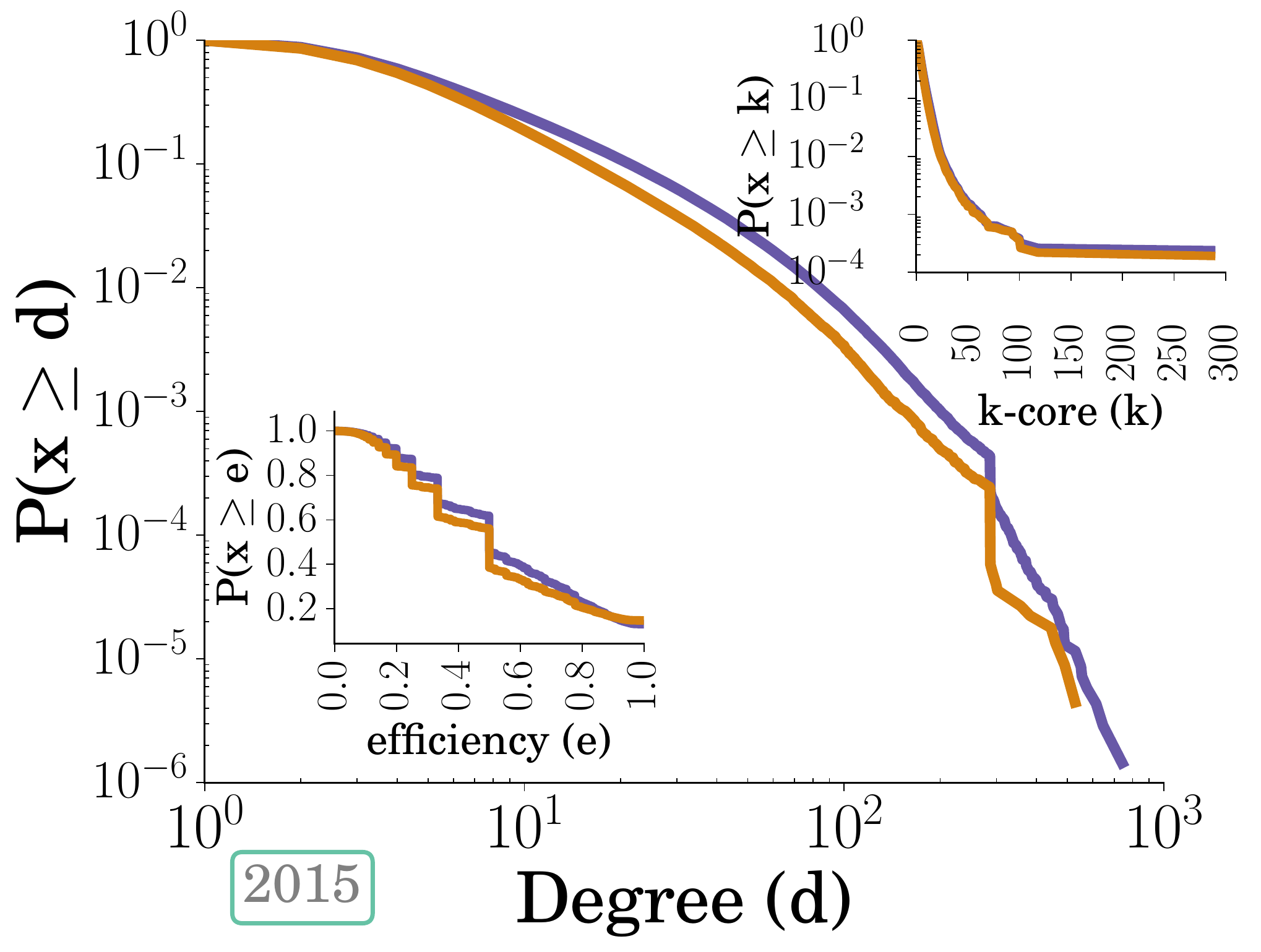}
\end{subfigure}
\caption{\textbf{Evolution of degree, \textit{k}-core and efficiency distributions over 6 decades}: Main figures show the degree distributions of male and female scientists. The top-right and bottom-left insets show the $k$-core and efficiency distributions, respectively. Each plot refers to one specific year and describes the structure of the network including all collaborations that occurred between the beginning of 1970 and the end of the given year.
As the cumulative network grows, the distributions grow fatter tails.
In the beginning (1970 and 1980), women tended to collaborate with fewer researchers (lower degree) and with researchers that were themselves less well connected (lower $k$-core) than men.
Women also tend to collaborate slightly more with colleagues that also collaborate with each other (lower efficiency).
}
\label{fig:kcore_degree_evo}
\end{figure*}

For structural analyses and later regressions analyses of gender and success we operationalize several concepts of network embeddedness. Node degree, the number of co-authors, is a measure of the \emph{size} of a researcher's ego network. Three measures offer insights into ego network properties. \emph{Cohesion} is the extent to which a network has evolved into a hierarchical structure of increasingly dense cores embedding into each other. Since the best operationalization is costly to implement \cite{moody2003structural} we use the $k$-core metric instead where $k$ is an ego's maximum number of co-authors that have at least $k$ neighbors themselves \cite{seidman1983network}.

Neither degree nor $k$-core tell if ego networks contain structural holes. Both the absence and the presence of such voids of connectivity are indispensible for the functioning of social networks. Closure, the absence of structural holes, is needed for trustful coordination while the presence of structural holes is accompanied by possibilities of brokerage, the reaping of advantages from tapping different pockets of information at multiple sides of the structural hole \cite{Burt2005}. We operationalize \emph{closure} through the clustering coefficient, the density of an ego network excluding ego \cite{watts_collective_1998}, and \emph{brokerage} using Burt's efficiency, the normalized number of co-authors minus their average degree within the ego network, excluding ties to ego \cite{Burt92}.

To also capture the dynamics of structural order and disorder -- or closure and brokerage -- we introduce two measures relating to team assembly \cite{guimera2005team}. \emph{Collaboration strength} is the median number of publications of ego's collaborations, and \emph{collaboration duration} is the median maximum publication year difference of ego's collaborations. If those scores are low, collaborations are less trustful, and brokerage is more pronounced.

\textbf{Structural gender disparities.}
Figure \ref{fig:kcore_degree_evo} depicts the growth of distributions of degree, $k$-core (top-right inset) and efficiency (bottom-left inset) for six points in cumulative time, distinguished by men and women.
The tails of the degree and $k$-core distributions reveal that collaboration at the macro level has been increasing over decades, regardless of gender. We also observe that, in earlier years, men have slightly broader degree and $k$-core distributions compared to women. As the total network grows and the number of women increases, women emerge with ego networks that are as sizable and cohesive as those of men.
With respect to efficiency, men tend to have slightly higher probabilities to act as bridges across structural holes. This is an intriguing result since previous work has shown that brokers tend to be more influential \cite{Ugander17042012,10.1086/421787,Burt92}.

\begin{table}[t]
\centering
\caption{\textbf{Cliff's \textit{d}-test to measure the distance between distributions.} Each value shows the $d$-statistic comparing degree, $k$-core and efficiency distributions for men and women
for networks cumulated up to the given year (cf. figure \ref{fig:kcore_degree_evo}).
Positive (negative) values indicate whether the distribution of men (women) is dominant.
The value of $d$ ranges from -1 (when every observation for women are greater that those of men) to 1 (when every observation for men are greater that those of women).
The differences between the distributions are significant but small for all years except the earlier ones when the network itself was small. In all significant cases, the distribution for men is dominant.
\textit{Note:} $^{*}p < 0.05$; $^{**}p < 0.01$; $^{***}p < 0.001$
\label{tab:cliff_test}}
\resizebox{\columnwidth}{!}{%
\def\colorrow{\rowcolor[gray]{0.75}}
\begin{tabular}{|l|l|l|l|l|l|l|}
\hline
&1970&1980&1990&2000&2010&2015
\\
\hline
\colorrow degree&0.000& 0.007***&0.023***& 0.064***&0.097***&0.069***
\\
\hline
k-core&0.000&0.007***&0.023***&0.063***&0.089***&0.051***
\\
\hline
\colorrow efficiency&-0.075&-0.028&0.002&0.027***&0.061***&0.074***
\\
\hline\end{tabular}
}
\end{table}

To quantify the comparison of these distributions for men and women, we use Cliff's $d$-test that measures the extent to which one distribution is statistically dominant over the other one \cite{cliff1993dominance}.
Table \ref{tab:cliff_test} gives the $d$-statistics
for degree, $k$-core and efficiency for six points in cumulative time.
We observe small but significant differences between the distributions. In all significant cases, the distribution for men is the dominant distribution -- i.e., men have larger and more cohesive networks, and they are more likely to be positioned at structural holes.

\begin{figure*}[t]
\centering
\begin{minipage}{.31\linewidth}
\centering
  \includegraphics[width=\textwidth]{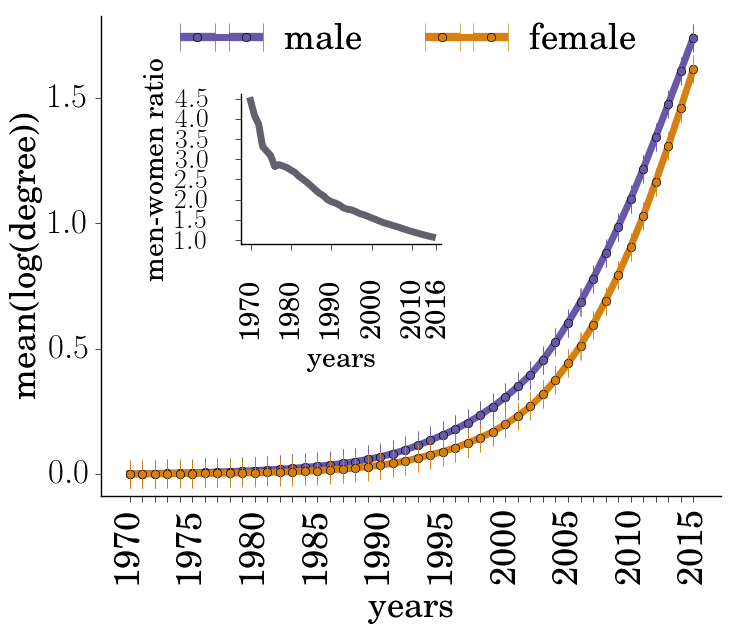}
\end{minipage}
\begin{minipage}{.31\linewidth}
\centering
  \includegraphics[width=\textwidth]{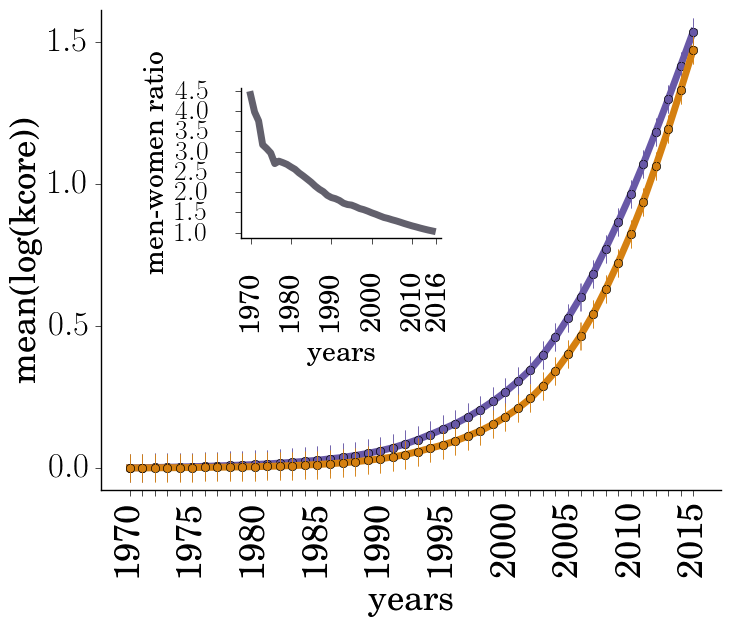}
\end{minipage}
\begin{minipage}{0.31\linewidth}
\centering
  \includegraphics[width=\textwidth]{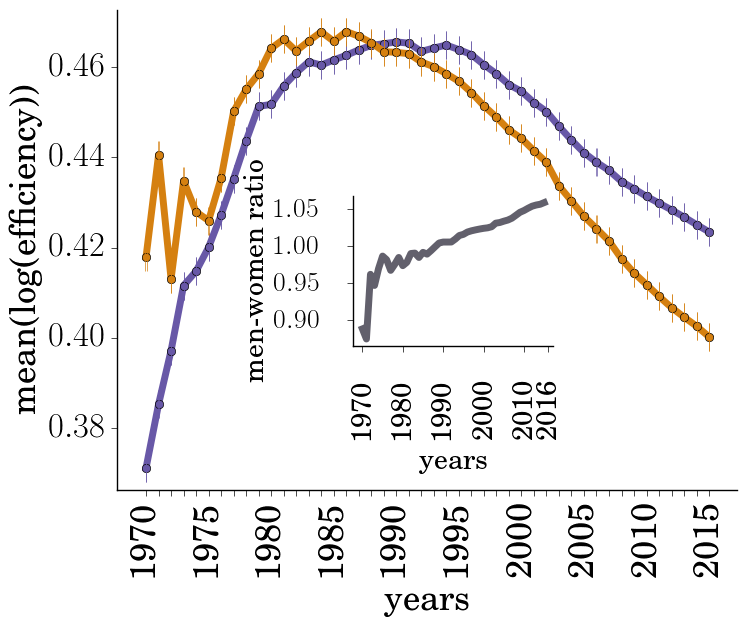}
\end{minipage}
\caption{\textbf{Changes of degree (left), \textit{k}-core (middle) and efficiency (right).} The main figures show the changes in means of log-transformed values over cumulative time. The insets show corresponding men-to-women ratios (ratios above (below) one indicate higher mean log-efficiency for men (women)). For degree and $k$-core men tend to have higher values, but the gap is decreasing over time.
The gender gap in efficiency shows three phases: In the first phase (1970--1982) women are stronger brokers than men (ratios are below 1). In the second phase (1983-1993) the average log-efficiencies are not distinguishable. In the third phase (1994-2015) men are stronger brokers.
\label{fig:evol_median}}
\end{figure*}

To quantify the change inherent to these distributions,
we study the mean of the log-transformed values and look at the men-to-women ratio over cumulative time.
Figure \ref{fig:evol_median} shows that
men tend to have larger and more cohesive networks at any time, though the gaps are decreasing.
Regarding brokerage, the gender gap closes until 1983, in 1989 men have higher log-efficiency for the first time, and by 1994 men are significantly stronger brokers on average.

\begin{table}[t]
\centering
\caption{\textbf{Association between collaboration features and gender.} Each model assesses the relationship between different collaboration features and gender ($male=0$, $female=1$) while controlling for the career age of scientists. Each cell gives the odds ratio from a logistic regression model that only uses a single collaboration feature to explain the gender of scientists in the collaboration network at the end of the given year. No significant effects are observed for early periods.
For periods up to more recent years, nodes with higher clustering coefficient, lower efficiency and lower collaboration duration are more likely to correspond to female scientists. Degree and $k$-core are significant but exhibit effect sizes close to 1.
We do not find any significant gender difference with respect to collaboration strength.
\textit{Note:} $^{*}p < 0.05$; $^{**}p < 0.01$; $^{***}p < 0.001$
\label{tab:gender_reg}}
\resizebox{\columnwidth}{!}{
\def\colorrow{\rowcolor[gray]{0.75}}
\begin{tabular}{|l|l|l|l|l|l|l|l|}
\hline
Period (1970--)&Clustering coefficient&Efficiency&$k$-Core&Degree& Collaboration duration&Collaboration strength&Sample size (female ratio)
\\[8pt]
\hline
\colorrow 1970& 0.0 (-0.0)&2.158 (0.627)&0.727 (-0.713) &0.667 (-0.967)& -~\tablefootnote[1]{~The z-score could not be computed because the corresponding value for all authors is equal to 1 and therefore standard deviation is equal to zero.} & - &228 (0.05)
\\[8pt]
\hline
1980& 1.669 (1.809)&0.81 (-0.754)&1.04 (1.414)&0.94 (-2.489)*& 0.922 (-2.18)* & 1.069 (0.726) &2,145 (0.09)
\\[8pt]
\hline
\colorrow  1990&1.916 (7.455)***&0.502 (-6.368)***&1.024 (2.602)**&0.969 (-5.798)***& 0.951 (-4.737)***& 1.008 (0.202)& 11,104 (0.13)
\\[8pt]
\hline
2000&1.412 (10.069)***&0.67 (-8.071)***&0.999 (-0.438)*&0.984 (-11.833)***& 0.951 (-11.013)***& 1.0 (-0.02)& 46,486 (0.16)
\\[8pt]
\hline
\colorrow  2010&1.649 (32.17)***&0.444 (-31.304)***&1.0 (-0.582)&0.99 (-25.03)***& 0.945 (-24.985)***&1.0112 (1.438) & 147,163 (0.2)
\\[8pt]
\hline
2015&1.818 (43.272)***&0.414 (-41.075)***&0.998 (-5.806)***&0.992 (-31.56)***&0.941 (-34.689)*** & 0.993 (-0.937)& 192,687 (0.21)
\\[8pt]
\hline
\end{tabular}
}
\end{table}

\textbf{Collaboration patterns across career ages.}
Although the results so far indicate that gender-specific differences in collaboration practices exist, other confounding factors, such as the career-age distribution of men and women or the computer-science specialties in which men and women are unequally embedded, may explain our results. To address this problem to some extent, we use multiple logistic regression models in which we use a single collaboration concept as the independent variable and gender as the dependent variable.

Diagnosing the relationship between position and gender requires accounting for dynamic effects. To explore the temporal stability of the bivariate relationships, we fit several models for increasing time periods (e.g., the model for the year 2000 is based on the cumulative collaboration network of all publications that have been published before or in 2000). To establish temporal comparability, we only study authors which are active in the final year of each period (e.g., the model for the year 2000 is based on those authors in the cumulative collaboration network which had published in 2000).

This reduces the sample size to the one given in the last column of figure \ref{tab:gender_reg}.

To further control for the career age of researchers, we replace a raw feature score $s$ by its corresponding career-age $z$-score separately for each period.
For example, for each scientist $i$ in a specific year, we measure how much her feature score at career age $\tau$, $s_i(\tau)$, deviates (in terms of standard deviation) from the average degree of scientists at the same career age:

\begin{equation}
\label{eq:zscore}
 z_i(\tau) = \frac{s_i(\tau) - \langle s(\tau) \rangle}{\sigma [s(\tau)]}
\end{equation}

Table \ref{tab:gender_reg} shows the odds ratio and $z$-statistics for each regression.
Before 1990 no significant effects can be observed.
For periods up to more recent years we find that scientists whose ego networks are more closed, contain fewer structural holes and are more short-lived are more likely to be female.
This statistical analysis confirms our earlier results that men and women do differ structurally, particularly regarding brokerage and closure, starting in the 90s.
The finding that women, on average, embed into networks with shorter collaboration duration may be interpreted to be in line with results by Zeng et al. \cite{Xiao2016} who found that women have a lower probability of repeating previous collaborations than men.
It should be noted that in all cases the coefficient of determination is close to zero, i.e. each feature alone can only explain a small proportion of variance in the response variable.

\textbf{Mixing of men and women.} \emph{Homophily}, the tendency to associate with similar others, is one of the fundamental factors that shape social ties \cite{moody2001race,Homophily_origin}. Homophilic behaviour combined with group size differences can limit minorities to stretch their overall degree \cite{karimi2017visibility}. Consequently, it can impact the opportunities afforded to minorities to access novel ideas and information.
Since we are interested in observing how homophily is changing over time, we analyze the collaborative behaviour of scientists within each year separately rather than looking at the accumulated collaboration network for each year.

\begin{figure}[t]
\centering
\begin{minipage}{0.47\linewidth}
\centering
  \includegraphics[width=\textwidth]{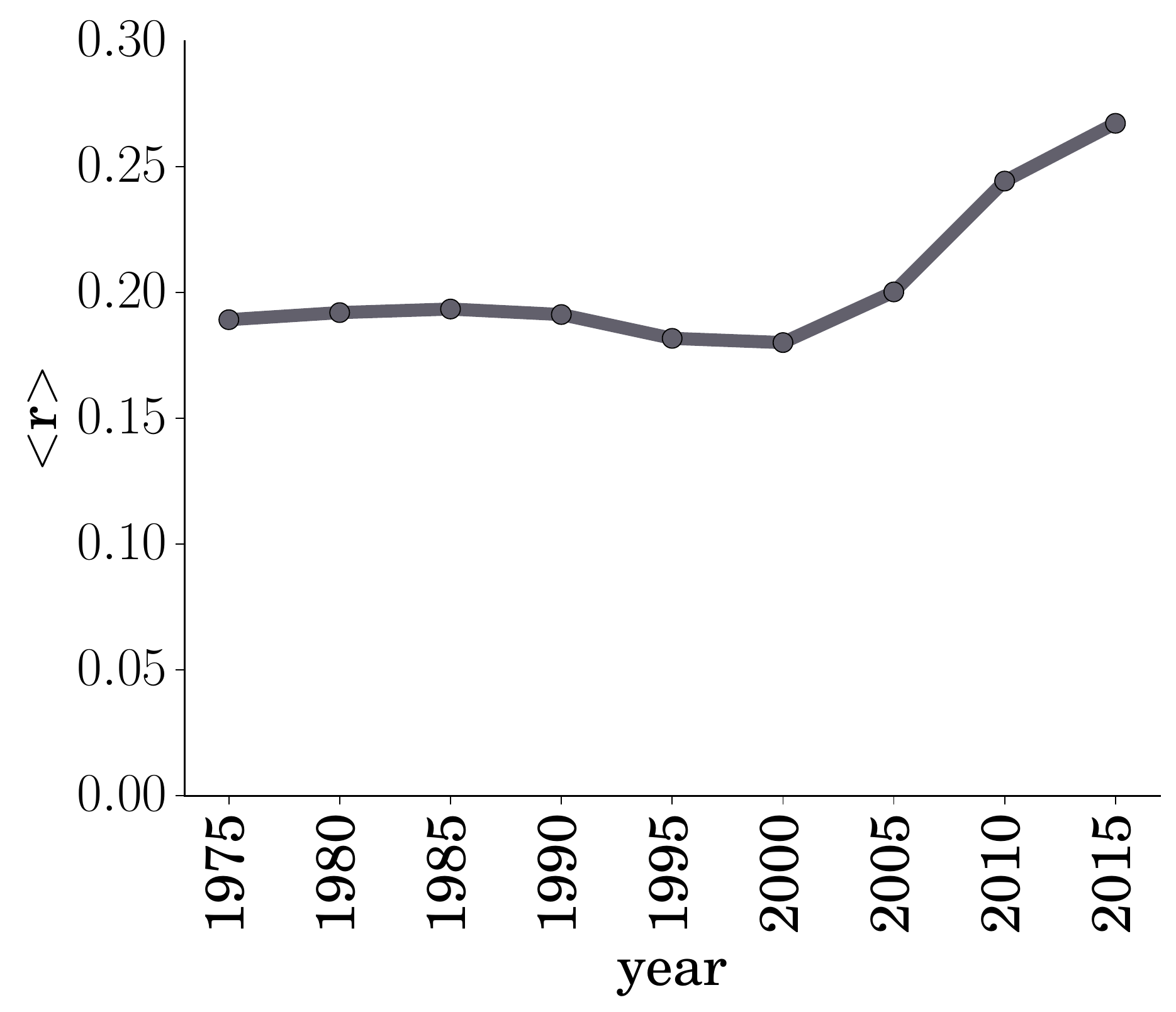}
\end{minipage}
\begin{minipage}{0.47\linewidth}
\centering
  \includegraphics[width=\textwidth]{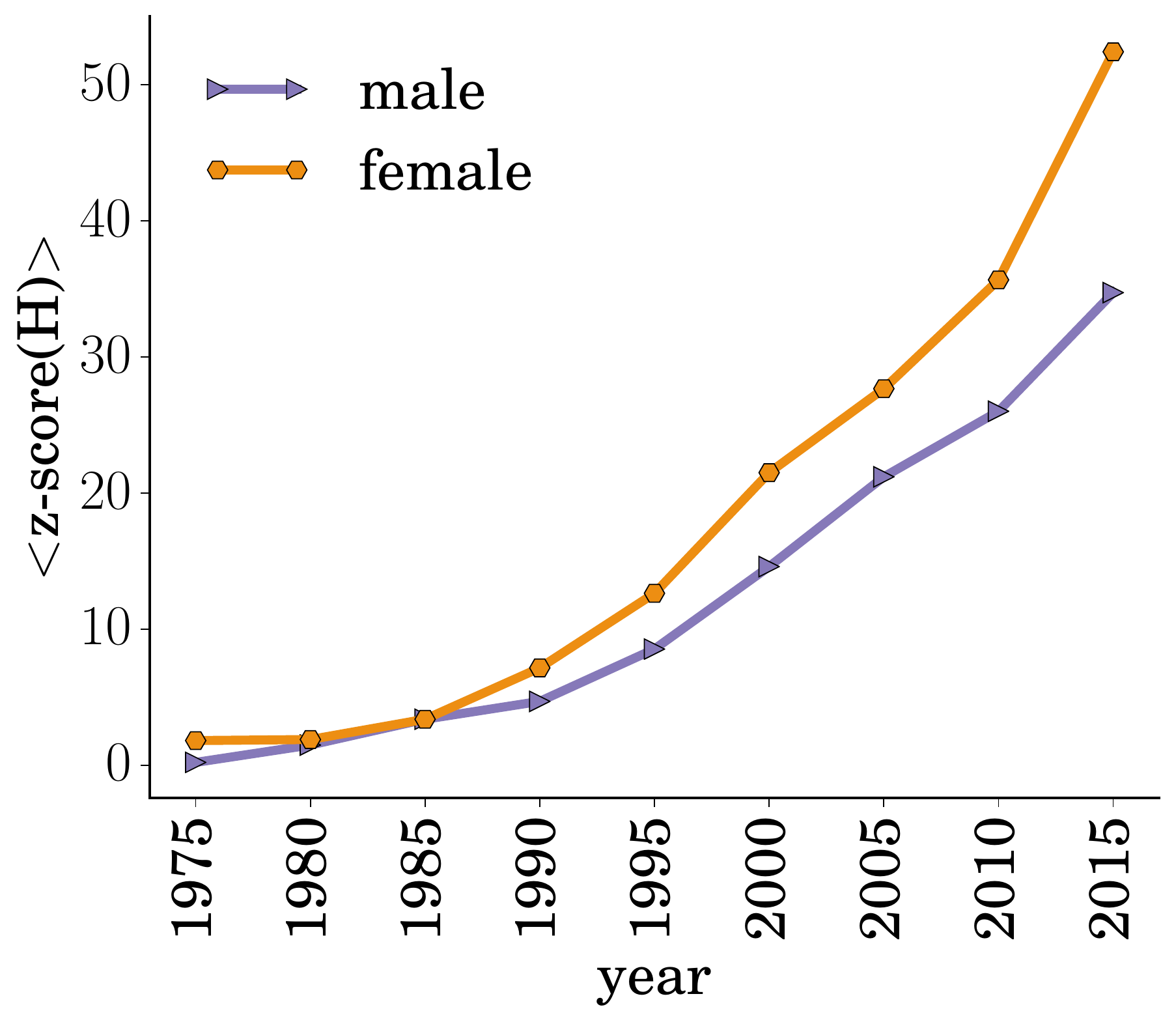}
\end{minipage}
\caption{\textbf{Gender assortativity and homophily.}
(Left) Newman gender assortativity $r$ computed for annual snapshots of the collaboration network.
Gender assortativity is stable until about 2000 and subsequently increases.
(Right) $z$-Score of homophily computed using equation \ref{eq:homophily} for annual snapshots and 100 instances of a corresponding null model (i.e. a network in which we reshuffle the links but keep the degree intact). $z$-Scores indicate the deviation (in terms of standard deviation) from the homophily we would expect in a randomized network. They are computed separately for men and women.
Homophily increases monotonically, women are more homophilic than men and the gap widens.
All curves are smoothed using a 5-year moving average.
\label{fig:gender_assor}
}
\end{figure}

To diagnose global changes of homophily, we use Newman's assortativity measure $r$ that captures the extent to which collaborative ties exist across gender ($r<0$) and among the same gender ($r>0$) compared to what we would expect from the node's degree \cite{newman2003mixing}. Figure \ref{fig:gender_assor} (left)
suggests that assortativity was relatively stable in the past but started to increase in 2000.

The increasing trend in gender assortativity requires a detailed analysis to uncover whether the increase is mainly produced by the behaviour of one group or both groups. To assess the homophily for each gender separately, we look at the proportion of links between women ($H_f$) and men ($H_m$):
\begin{equation} \label{eq:homophily}
 \begin{split}
     & H_m = \frac{E_{m,m}}{E_{m,m}+E_{f,m}}\\
     & H_f = \frac{E_{f,f}}{E_{f,f}+E_{f,m}}
 \end{split}
 \end{equation}
Here $E_{m,m}$ refers to male-to-male edges, $E_{f,f}$ to female-to-female edges, and $E_{f,m}$ to female-to-male edges.
For example, $H_f = 1$ means that women only collaborate with women.

To assess the significance of the observed mixing pattern, we compare the observation to null models in which we keep the network size and the degree of the nodes intact and reshuffle the edges. Using this model we generate 100 synthetic networks for each yearly snapshot of our empirically observed co-authorship network. The synthetic networks represent random baselines that are expected if men and women are gender-blind during co-author selection.
As a last step, we compute the mean and standard deviation of male and female homophily and the report the corresponding $z$-score.

Figure \ref{fig:gender_assor} (right) shows
how many standard deviations the empirical homophily deviates from the expectation if the interactions would not be impacted by gender. We again see that the homophilic behaviour of men and women is increasing over time. However, the homophilic behaviour of women exceeds the expectation more than those of men.

Note that our baseline model assumes that every computer scientist can in theory collaborate with any other computer scientist. In reality subfields and specialties constrain who could collaborate with whom. If women are a minority that focuses on selected topical areas (e.g., Human Computer Interaction), then we would observe higher homophily for women than expected from our baseline model, assuming that collaborations within subfields are more likely than across subfields. That means, while our work shows that women tend to collaborate more with other women than expected, we do not answer the question why this is happening. Gender is one possible explanation, but also the gender composition of certain subfields will play a role. Therefore, whether the observed homophily is the result of authors' choices (choice homophily) or emergent structures (induced homophily) requires a deeper investigation that we leave for future works. \cite{Homophily_origin,shalizi_homophily_2011}

\subsection{Success}

Here, we aim to understand the relationship between collaboration patterns, gender and scientific success. Specifically, we seek to answer \emph{which collaboration patterns are related with scientific success and if these patterns are similar for male and female scientists}.
To quantify scientific success, our dependent variable, we use two common measures: \emph{citation impact}, the raw number of citations an author has accumulated up to a given year, and the $h$-index, the number of an author's publications that have accumulated at least $h$ citations \cite{hirsch2005index}.
While the number of citations can be driven by single high-impact papers, the $h$-index combines the assessment of both quantity (number of papers) and quality (number of citations). A scientist needs to produce a high number of high quality papers in order to obtain a high $h$-index.

We create two different regression models that describe the relationship between the collaborative behaviour of scientists and their success.

The first model (\emph{ego model}) relies on
the ego-centric properties of a node defined in the previous subsection.
Because of a high correlation between degree and $k$-core (Pearson correlation of $0.75$ with $p < 0.001$), we do not use $k$-core in our model to avoid multicollinearity.
The second model (\emph{1-hop model}) extends the ego model by including information about a node's median neighbourhood structure.

Moreover, the academic system naturally changes over time (e.g., with respect to size, number of relevant venues, publication and citation practices). Therefore comparing scientists that started their career in different decades may confound our results. To control for this effect, we add the starting decade of an author's career to our model. 
To study the effect of gender in collaboration and on success, we include gender as an interaction term in our models.

\begin{table}[t]
\centering
\caption{\textbf{Sample size for regression of success.} Beside the number of authors we also list the number of observations since we have multiple observations per author (one for each year in which they were active).
\label{tab:reg_stats}}
\def\colorrow{\rowcolor[gray]{0.75}}
\begin{tabular}{|l|l|l|l|}
\hline
&Men&Women&Total
\\
\hline
Number of authors & 72,076 & 13,746& 85,822
\\
\hline
Number of observations &734,474 & 131,194&865,668
\\
\hline\end{tabular}
\end{table}

The population of scientists is restricted to those with careers of at least 10 years and at least 5 publications.
This way we focus only on people who have decided to pursue an academic career.
For each scientist we record her collaborative features for all stages of her academic career,
i.e. our panel data consists of multiple observations (at least 5) for each author, one for each career age.
Furthermore, we ignore the first 5 career ages to give authors enough time to accumulate citations. Table \ref{tab:reg_stats} shows the size of our panel.

To account for within-subject correlation and unbalanced observations for subjects (e.g., missing observations), we use the General Estimation Equation (GEE) regression model \cite{citeulike:849124} with an \emph{exchangeable correlation structure}.
This structure meets our cumulative research design by assuming that the correlations between features for the same author at different career ages are stationary..
We fit the GEE model with a Gaussian distribution and the identity link function to the data. To assess the goodness of the fit we use the marginal $R^2$ which is an extension of $R^2$ statistics for GEE models \cite{SIM:SIM486}. Similar to $R^2$, marginal $R^2$ can be interpreted as the proportion of variance in the response variable explained by the fitted model.

We consider a scientist as successful if she has a higher citation impact or $h$-index than an average scientists in the same career age. Therefore we again use equation \ref{eq:zscore} to compute the age-specific $z$-scores for the number of citations and the $h$-index.
Since the $z$-scores of our dependent variables are skewed, we use of log of the $z$-scores instead.
The independent variables are transformed into $z$-scores but not logged. 
Therefore, the coefficients quantify the association between above-average collaboration features and success.

\begin{table}[ht!]
\centering
\caption{\textbf{GEE model for citation impact.}
Odd ratios of coefficients are given for the number of citations as the dependent variable. Values in brackets give $z$-statistics for the coefficients.
The ego model shows that degree, collaboration duration and collaboration strength are sizeably and positively related to scientific success. Efficiency has a small positive but significant effect.
The 1-hop neighbourhood model confirms these observations and finds that median number of citations as well as career age of alters significantly add to ego's success while clustering coefficient of alters has a negative effect.
There is no gender effect.
\textit{Note:} $^{*}p < 0.05$; $^{**}p < 0.01$; $^{***}p < 0.001$
\label{tab:reg-median-ncit}
}
\def\colorrow{\rowcolor[gray]{0.75}}
\resizebox{\columnwidth}{!}{%
\begin{tabular}{|>{\LARGE}l|>{\LARGE}l|>{\LARGE}l|>{\LARGE}l|>{\LARGE}l|}
\hline
&ego model& \begin{tabular}[x]{@{}c@{}}ego model\\+interactions\end{tabular}& 1-hop model&\begin{tabular}[x]{@{}c@{}}1-hop model\\+interactions\end{tabular}
\\ [10pt]
\hline
\colorrow  Intercept&1.682(125.699)***&1.681(125.098)***&1.700(133.419)***&1.700(132.788)***
\\[8pt]
\hline
 gender(reference=men)[women]&&1.004(1.518)&&1.002(0.783)
\\[8pt]
\hline
 \colorrow clustering&0.999(-1.234)&0.999(-1.213)&0.999(-2.098)&0.999(-1.769)
\\[8pt]
\hline
 clustering*gender&&1.000(0.211)&&1.000(-0.297)
\\[8pt]
\hline
\colorrow degree&1.128(69.524)***&1.127(63.104)***&1.122(58.912)***&1.120(53.123)***
\\[8pt]
\hline
degree*gender&&1.007(1.364)&&1.008(1.659)
\\[8pt]
\hline
\colorrow efficiency&1.004(7.467)***&1.004(6.88)***&1.005(7.811)***&1.005(7.95)***
\\[8pt]
\hline
efficiency*gender&&0.999(-0.499)&&0.998(-1.263)
\\[8pt]
\hline
\colorrow median collaboration duration&1.021(57.046)***&1.021(52.313)***&1.015(34.412)***&1.015(31.314)***
\\[8pt]
\hline
median collaboration duration*gender&&1.000(0.048)&&1.000(-0.177)
\\[8pt]
\hline
\colorrow median collaboration strength&1.007(10.833)***&1.007(9.732)***&1.005(8.445)***&1.005(8.039)***
\\[8pt]
\hline
median collaboration strength*gender&&0.998(-1.092)&&0.999(-1.051)
\\[8pt]
\hline
\colorrow neighbours median age&&&1.004(5.473)***&1.004(4.872)***
\\[8pt]
\hline
neighbours median age*gender&&&&1.000(-0.007)
\\[8pt]
\hline
\colorrow neighbours median clustering&&&0.998(-5.666)***&0.997(-5.403)***
\\[8pt]
\hline
neighbours median clustering*gender&&&&1.001(0.894)
\\[8pt]
\hline
\colorrow neighbours median degree&&&1.005(1.613)&1.006(1.71)
\\[8pt]
\hline
neighbours median degree*gender&&&&0.993(-1.701)
\\[8pt]
\hline
\colorrow neighbours median n citations&&&1.036(6.376)***&1.036(5.452)***
\\[8pt]
\hline
neighbours median n citations*gender&&&&1.000(0.036)
\\[8pt]
\hline
\colorrow 
&&&&\\[8pt]
\hline
Marginal $R^2$&0.217&0.217&0.296&0.296
\\[8pt]
\hline\end{tabular}
}
\end{table}

\begin{table}[ht!]
\centering
\caption{\textbf{GEE model for \textit{h}-index.}
Odd ratios of coefficients are given for the $h$-index as the dependent variable. Values in brackets give $z$-statistics for the coefficients.
The ego model shows that degree, collaboration duration, efficiency and collaboration strength are sizeably related to scientific success. Clustering coefficient has a small but significant effect.
The 1-hop neighbourhood model confirms these observations and finds that the median career age sizeably and the number of citations as well as the degree of alters significantly add to ego's success.
There is no gender effect.
\textit{Note:} $^{*}p < 0.05$; $^{**}p < 0.01$; $^{***}p < 0.001$
\label{tab:reg-median-hindex}}
\def\colorrow{\rowcolor[gray]{0.75}}
\resizebox{\columnwidth}{!}{%
\begin{tabular}{|>{\LARGE}l|>{\LARGE}l|>{\LARGE}l|>{\LARGE}l|>{\LARGE}l|}
\hline
&ego model& \begin{tabular}[x]{@{}c@{}}ego model\\+interactions\end{tabular}& 1-hop model&\begin{tabular}[x]{@{}c@{}}1-hop model\\+interactions\end{tabular}
\\[10pt]
\hline
\colorrow  Intercept&1.971(100.784)***&1.968(100.382)***&2.009(107.311)***&2.008(107.022)***
\\[8pt]
\hline
  gender(reference=men)[women] &&1.016(5.396)***&&1.012(4.485)***
\\[8pt]
\hline
  \colorrow clustering&1.005(7.335)***&1.005(6.676)***&1.005(6.718)***&1.005(6.215)***
\\[8pt]
\hline
clustering*gender&&1.000(0.026)&&0.999(-0.404)
\\[8pt]
\hline
\colorrow degree&1.172(96.373)***&1.172(87.289)***&1.157(78.669)***&1.156(71.131)***
\\[8pt]
\hline
degree*gender&&1.002(0.428)&&1.005(0.993)
\\[8pt]
\hline
\colorrow efficiency&1.017(23.733)***&1.016(21.355)***&1.021(26.531)***&1.021(25.288)***
\\[8pt]
\hline
efficiency*gender&&0.999(-0.331)&&0.997(-1.326)
\\[8pt]
\hline
\colorrow median collaboration duration&1.032(66.816)***&1.032(61.709)***&1.022(41.479)***&1.022(38.399)***
\\[8pt]
\hline
median collaboration duration*gender&&0.998(-1.832)&&0.998(-1.423)
\\[8pt]
\hline
\colorrow median collaboration strength&1.013(10.364)***&1.013(8.727)***&1.009(8.303)***&1.009(7.256)***
\\[8pt]
\hline
median collaboration strength*gender&&0.998(-0.965)&&0.998(-0.788)
\\[8pt]
\hline
\colorrow neighbours median age&&&1.016(16.516)***&1.016(15.386)***
\\[8pt]
\hline
neighbours median age*gender&&&&0.995(-2.137)
\\[8pt]
\hline
\colorrow neighbours median clustering&&&0.999(-1.441)&0.999(-1.437)
\\[8pt]
\hline
neighbours median clustering*gender&&&&1.000(0.087)
\\[8pt]
\hline
\colorrow neighbours median degree&&&1.018(6.28)***&1.019(5.883)***
\\[8pt]
\hline
neighbours median degree*gender&&&&0.991(-2.134)
\\[8pt]
\hline
\colorrow neighbours median n citations&&&1.035(6.441)***&1.034(5.501)***
\\[8pt]
\hline
neighbours median n citations*gender&&&&1.004(0.513)
\\[8pt]
\hline
\colorrow 
&&&&\\[8pt]
\hline
Marginal $R^2$&0.220&0.220&0.298&0.298
\\[8pt]
\hline\end{tabular}
}
\end{table}

Tables \ref{tab:reg-median-ncit} and \ref{tab:reg-median-hindex} report odd ratios and size effects for the number of citations and the $h$-index, respectively, as proxies for success.
All four models (the ego and 1-hop models for citation impact and $h$-index) agree that embedding into large enduring networks with some repetition of collaborations is the primary explanation of academic success. Structural closure is a significant predictor in the $h$-index models.
Brokerage, however, the tapping of various information resources, is also a significant predictor of success, even a strong one when success is measured through the $h$-index. Interestingly, in the latter case, closure also turns significant. 
Much in line with the existing literature \cite{guimera2005team,palla2007quantifying} this means that trustful relations are not an option but a requirement for authors and fields to thrive.
Successful scientists keep reproducing a large network of core collaborators while simultaneously adding new collaborators from a variety of social circles.
While long-lasting research partnerships can lead to collaborations that increase success through increased productivity \cite{petersen_tie}, new collaborators and brokerage can increase visibility within the community and make a researcher more influential \cite{Ugander17042012,10.1086/421787}.

In addition to the effects of ego-centric features, the 1-hop models demonstrate that collaborating with successful and senior scientists is beneficial for a researcher, especially in the $h$-index models. In that case, when success is also assessed in terms of productivity, collaborating with highly-connected scientists is also beneficial. This is probably the effect that teams of junior and senior researchers can produce outputs of quantity and quality \cite{Wuchty1036}.
Given that, in the ego model, creating trustful relations (enduring and strong) is a stronger predictor of success than brokerage, one may expect that collaborations with strong brokers may be beneficial. The observation that ties to co-authors with highly closed networks have a negative effect on success may be considered as supporting evidence for this conjecture.

Finally, our analysis shows that no significant gender-specific differences exist in how collaboration patterns impact success, since no interactions between gender and collaboration patterns can be found.
This is evidence that \emph{successful male and female scientists exhibit the same collaborative behaviour} and that \emph{no differences exist in which collaboration patterns may explain the success of men and women in computer science}.

While the same collaboration patterns explain the success of male and female scientists, our previous analysis (see table \ref{tab:gender_reg}) revealed that men and women do embed into significantly different ego networks.
Networks of female researchers are significantly smaller, more closed, more devoid of structural holes and -- on the median -- more short-lived, while men take roles as explorers of large spaces who maintain trustful relations on the long run. Male collaborative behaviour is the one associated with success in academia.
This suggests that women are on average less likely to adapt the collaborative behaviour that is related to success. However, those women who do become successful computer scientists show the same collaboration patterns as their successful male colleagues.

Interestingly, gender has a minuscule but significant effect on the $h$-index but not on the number of citations in the 1-hop models (tables \ref{tab:reg-median-ncit} and \ref{tab:reg-median-hindex}).
Also note that the regression models only explain 20--30\% of the variance, i.e. our purely structural approach misses central aspects of the research practice.

\section{Discussion}

A potential gender gap in academia, especially in STEM fields, has been a great concern over the past decades and many studies have tried to quantify the extent to which gender inequalities are present in science.

In this study, we have focused on the collaborative behaviour of scientists in one entire scientific field, computer science, a densifying field \cite{yang2012structure} with a collaborative style \cite{Wuchty1036} and a wide gender gap \cite{Leslie262}.
We have taken a career approach and analyzed how male and female scientists differ in their dropout rates, their productivity, their tendency to associate with same-sex researchers as well as their collaborative behaviour and its relation with success in academia.

We find that the dropout rate of women in computer science is consistently higher than of men. Women also have smaller probabilities to continue after their first publication year, enter early- and mid-career stages and become senior researchers.
Controlling for this difference in careers is our solution to the \emph{productivity puzzle} that women have a smaller publication output than men. The solution is found in a strong correlation between the \emph{productivity gap} and what we call the \emph{seniority gap}. Put simply, men are more productive on average because they have a larger fraction of senior authors. We solve this puzzle without the need to refer to exogenous factors or use other sources but bibliographic data \cite{Duch2012,wenneras2001nepotism,bentley_2003,Stack2004,prozesky2008career,knobloch2013matilda}.

There is no gender effect regarding success. Women are more likely to have a larger $h$-index than men, but the gender variable does not interact with any of the collaboration features. Structural insights from regressions of success on collaborative patterns over researchers' careers are more revealing. They show that network closure \emph{and} network brokerage are co-determinants of citation impact or the $h$-index.
Successful scientists embed in large networks and build trustful relationships through repeating collaborations throughout their careers \cite{petersen_tie}. But, at the same time, successful scientists also bridge structural holes to exploit various knowledge resources and stay innovative \cite{Ugander17042012,10.1086/421787}, making brokerage and closure a true duality \cite{Burt2005,abbasi_identifying_2011}. This resonates with theories in the sociology of science that tradition and innovation is the ``essential tension'' in scientific research \cite{kuhn_essential_1959}, or in organization science that the exploration of new possibilities necessarily complements the exploitation of old certainties \cite{march_exploration_1991}.

Interestingly, our temporal classification of gender on collaborative behaviour demonstrates that male and female researchers do embed into different ego networks on average. ``Female'' networks are significantly smaller, much more closed (clustered), contain fewer brokerage opportunities and are more short-lived (regarding median collaboration durations) than those of men. Controlling for dynamic effects and career ages, there is a division of labour in computer science: Women tend to take care of network closure and gather knowledge in tightly-knit communities while men tend to hunt for innovations across structural holes.
Related disparities have been observed in managerial networks where women are more successful with a small network of interconnected contacts \cite{burt_gender_1998} or where trustful relations and group reproduction have been found to increase with the presence of women \cite{westermann_gender_2005}.

While the difference in network structure for men and women opens up future venues of research into potential gender gaps in science, it is not in conflict with our results that there are no gender-specific differences in how collaboration patterns impact success.

The suggested solution to the seeming paradox is that women are on average less likely to adapt the collaboration patterns that are related with success. However, those women who become successful computer scientists exhibit the same collaborative behaviour as their successful male colleagues.

Regarding the mixing of men and women over time we find that gender homophily has been increasing ever since. In particular, homophily among women is higher than among men when controlling for network topology and size.

\textbf{Limitations.}
Our work does not allow to answer causal questions, such as if certain collaboration strategies (e.g., repetitive collaborations or bringing people from different communities together) lead to success or if the observed patterns are a consequence of success.

It is very likely that these relationships are not unidirectionally causal, but mediated by an unobserved variable, the skills and knowledge of a scientist \cite{Sinatraaaf5239}. 
Although our statistical models controlled for different factors such as career age, our work is limited to characteristics that are measurable and observable in our social network data.
Since academic fields are dualities of social networks and cultural domains \cite{white_identity_2008}, future studies should incorporate the actual content of fields, for example by detecting and adding latent variables representing subfields to regression models \cite{shalizi_homophily_2011}.

Finally, our results are limited to the non-Asian part of the computer science community since we excluded Asian names to avoid low precision in the gender inference task.

\textbf{Contributions.}
To our best knowledge, this is the first study that analyzes the productivity, dropouts, collaboration practices and success of male and female scientists in one entire scientific field over time. We hope that this work enhances our understanding of gender-specific differences in collaborative academic behaviour, how these differences change over time and how collaboration practices are related with success.
For future work it would be interesting to extend this analysis to more academic fields, explore disparities across ethnic groups and improve gender inference methods for Asian names.

\section*{\large Acknowledgements}
We thank Markus Strohmaier and our three reviewers for their invaluable comments on drafts of the paper.

\bibliographystyle{ws-acs}
\bibliography{ws-acs}

\begin{thebibliography}{10}
\providecommand{\urlprefix}{}
\expandafter\ifx\csname urlstyle\endcsname\relax
  \providecommand{\doi}[1]{doi:\discretionary{}{}{}#1}\else
  \providecommand{\doi}{doi:\discretionary{}{}{}\begingroup
  \urlstyle{rm}\Url}\fi

\bibitem{abbasi_identifying_2011}
Abbasi, A., Altmann, J., and Hossain, L., Identifying the effects of
  co-authorship networks on the performance of scholars: {A} correlation and
  regression analysis of performance measures and social network analysis
  measures, \emph{Journal of Informetrics} \textbf{5} (2011) 594--607.

\bibitem{ASI:ASI21486}
Aksnes, D.~W., Rorstad, K., Piro, F., and Sivertsen, G., Are female researchers
  less cited? {A} large-scale study of {N}orwegian scientists, \emph{Journal of
  the American Society for Information Science and Technology} \textbf{62}
  (2011) 628--636.

\bibitem{Bayer1977}
Bayer, A.~E. and Dutton, J.~E., Career age and research-professional activities
  of academic scientists: Tests of alternative nonlinear models and some
  implications for higher education faculty policies, \emph{The Journal of
  Higher Education} \textbf{48} (1977) 259--282.

\bibitem{bentley_2003}
Bentley, J.~T. and Adamson, R., Gender differences in the careers of academic
  scientists and engineers: A literature review, Special report, National
  Science Foundation (2003).

\bibitem{seidman1983network}
B.Seidman, S., Network structure and minimum degree, \emph{Social Networks}
  \textbf{5} (1983) 269--287.

\bibitem{Burt92}
Burt, R.~S., \emph{Structural holes: The social structure of competition}
  (Harvard University Press, 1995).

\bibitem{burt_gender_1998}
Burt, R.~S., The gender of social capital, \emph{Rationality and Society}
  \textbf{10} (1998) 5--46.

\bibitem{10.1086/421787}
Burt, R.~S., Structural holes and good ideas, \emph{{American Journal of
  Sociology}} \textbf{110} (2004) 349--399.

\bibitem{Burt2005}
Burt, R.~S., \emph{Brokerage \& Closure: An Introduction to Social Capital}
  (Oxford University Press, 2005).

\bibitem{cliff1993dominance}
Cliff, N., Dominance statistics: Ordinal analyses to answer ordinal questions,
  \emph{Psychological Bulletin} \textbf{114} (1993) 494.

\bibitem{cole_1984}
Cole, J.~R. and Zuckerman, H., The productivity puzzle: {P}ersistence and
  changes in patterns of publication of men and women scientists,
  \emph{Advances in Motivation and Achievements} \textbf{2} (1984) 17–--256.

\bibitem{Ding2006}
Ding, W.~W., Murray, F., and Stuart, T.~E., Gender differences in patenting in
  the academic life sciences, \emph{Science} \textbf{313} (2006) 665--667.

\bibitem{Duch2012}
Duch, J., Zeng, X. H.~T., Sales-Pardo, M., Radicchi, F., Otis, S., Woodruff,
  T.~K., and Nunes~Amaral, L.~A., The possible role of resource requirements
  and academic career-choice risk on gender differences in publication rate and
  impact, \emph{PLOS ONE} \textbf{7} (2012) 1--11.

\bibitem{gallivan2015co}
Gallivan, M. and Ahuja, M., Co-authorship, homophily, and scholarly influence
  in information systems research, \emph{Journal of the Association for
  Information Systems} \textbf{16} (2015) 980--1015.

\bibitem{guimera2005team}
Guimer\`{a}, R., Uzzi, B., Spiro, J., and Amaral, L. A.~N., Team assembly
  mechanisms determine collaboration network structure and team performance,
  \emph{Science} \textbf{308} (2005) 697--702.

\bibitem{hirsch2005index}
Hirsch, J.~E., An index to quantify an individual's scientific research output,
  \emph{Proceedings of the National Academy of Sciences of the United States of
  America} \textbf{102} (2005) 16569.

\bibitem{Holden2001}
Holden, C., General contentment masks gender gap in first {AAAS} salary and job
  survey, \emph{Science} \textbf{294} (2001) 396--411.

\bibitem{Kaatz2014371}
Kaatz, A., Gutierrez, B., and Carnes, M., Threats to objectivity in peer
  review: the case of gender, \emph{Trends in Pharmacological Sciences}
  \textbf{35} (2014) 371--373.

\bibitem{karimi2017visibility}
{Karimi}, F., {G{\'e}nois}, M., {Wagner}, C., {Singer}, P., and {Strohmaier},
  M., {Visibility of minorities in social networks}, \emph{ArXiv e-prints}
  \textbf{1702.00150} (2017).

\bibitem{Karimi:2016}
Karimi, F., Wagner, C., Lemmerich, F., Jadidi, M., and Strohmaier, M.,
  Inferring gender from names on the web: A comparative evaluation of gender
  detection methods, in \emph{Proceedings of the 25th International Conference
  Companion on World Wide Web} (International World Wide Web Conferences
  Steering Committee, 2016), pp. 53--54.

\bibitem{knobloch2013matilda}
Knobloch-Westerwick, S., Glynn, C.~J., and Huge, M., The {M}atilda {E}ffect in
  science communication: {A}n experiment on gender bias in publication quality
  perceptions and collaboration interest, \emph{Science Communication}
  \textbf{35} (2013) 603--625.

\bibitem{Homophily_origin}
Kossinets, G. and Watts, D.~J., Origins of homophily in an evolving social
  network, \emph{{American Journal of Sociology}} \textbf{115} (2009) 405--450.

\bibitem{kuhn_essential_1959}
Kuhn, T.~S., The essential tension: Tradition and innovation in scientific
  research, in \emph{{The Third (1959) University of Utah Research Conference
  on the Identification of Scientific Talent}} (University of Utah, 1959), pp.
  162--174.

\bibitem{kyvik1996child}
Kyvik, S. and Teigen, M., Child care, research collaboration, and gender
  differences in scientific productivity, \emph{Science, Technology \& Human
  Values} \textbf{21} (1996) 54--71.

\bibitem{Cassidy2013}
Larivi\`{e}re, V., Ni, C., Gingras, Y., Cronin, B., and Sugimoto, C.~R.,
  {Bibliometrics: Global gender disparities in science}, \emph{Nature}
  \textbf{504} (2013) 211--213.

\bibitem{Leahey01122006}
Leahey, E., Gender differences in productivity: Research specialization as a
  missing link, \emph{Gender \& Society} \textbf{20} (2006) 754--780.

\bibitem{Lehman1954}
Lehman, H.~C., Men's creative production rate at different ages and in
  different countries, \emph{The Scientific Monthly} \textbf{78} (1954)
  321--326.

\bibitem{Leslie262}
Leslie, S.-J., Cimpian, A., Meyer, M., and Freeland, E., Expectations of
  brilliance underlie gender distributions across academic disciplines,
  \emph{Science} \textbf{347} (2015) 262--265.

\bibitem{DBLP:journals/pvldb/Ley09}
Ley, M., {DBLP} - some lessons learned, \emph{Proc. {VLDB} {E}ndowment}
  \textbf{2} (2009) 1493--1500.

\bibitem{Ley2008}
Ley, T.~J. and Hamilton, B.~H., The gender gap in {NIH} grant applications,
  \emph{Science} \textbf{322} (2008) 1472--1474.

\bibitem{citeulike:849124}
Liang, K.-Y. and Zeger, S.~L., {Longitudinal data analysis using generalized
  linear models}, \emph{Biometrika} \textbf{73} (1986) 13--22.

\bibitem{Macaluso2016}
Macaluso, B., Larivi\`{e}re, V., Sugimoto, T., and Sugimoto, C., Is science
  built on the shoulders of women? {A} study of gender differences in
  contributorship, \emph{Academic Medicine} \textbf{91} (2016) 1136--1142.

\bibitem{march_exploration_1991}
March, J.~G., Exploration and exploitation in organizational learning,
  \emph{Organization Science} \textbf{2} (1991) 71--87.

\bibitem{ECIN:ECIN68}
Mcdowell, J.~M. and Smith, J.~K., The effect of gender-sorting on propensity to
  coauthor: Implications for academic promotion, \emph{Economic Inquiry}
  \textbf{30} (1992) 68--82.

\bibitem{moody2001race}
Moody, J., Race, school integration, and friendship segregation in {America},
  \emph{{American Journal of Sociology}} \textbf{107} (2001) 679--716.

\bibitem{moody2004structure}
Moody, J., The structure of a social science collaboration network:
  Disciplinary cohesion from 1963 to 1999, \emph{American Sociological review}
  \textbf{69} (2004) 213--238.

\bibitem{moody2003structural}
Moody, J. and White, D.~R., Structural cohesion and embeddedness: A
  hierarchical concept of social groups, \emph{American Sociological Review}
  \textbf{68} (2003) 103--127.

\bibitem{Moss2012}
Moss-Racusin, C.~A., Dovidio, J.~F., Brescoll, V.~L., Graham, M.~J., and
  Handelsman, J., {Science faculty's subtle gender biases favor male students},
  \emph{Proceedings of the National Academy of Sciences of the United States of
  America} \textbf{109} (2012) 16474--16479.

\bibitem{doi:10.1027/2151-2604/a000103}
Mutz, R., Bornmann, L., and Daniel, H.-D., Does gender matter in grant peer
  review?, \emph{Zeitschrift f\"{u}r Psychologie} \textbf{220} (2012) 121--129.

\bibitem{newman2003mixing}
Newman, M.~E., Mixing patterns in networks, \emph{Physical Review E}
  \textbf{67} (2003) 026126.

\bibitem{palla2007quantifying}
Palla, G., Barab{\'a}si, A.-L., and Vicsek, T., Quantifying social group
  evolution, \emph{Nature} \textbf{446} (2007) 664--667.

\bibitem{pell1996fixing}
Pell, A.~N., Fixing the leaky pipeline: {W}omen scientists in academia,
  \emph{Journal of animal science} \textbf{74} (1996) 2843--2848.

\bibitem{petersen_tie}
Petersen, A.~M., {Quantifying the impact of weak, strong, and super ties in
  scientific careers}, \emph{Proceedings of the National Academy of Sciences of
  the United States of America} \textbf{112} (2015) E4671--E4680.

\bibitem{petersen2014reputation}
Petersen, A.~M., Fortunato, S., Pan, R.~K., Kaski, K., Penner, O., Rungi, A.,
  Riccaboni, M., Stanley, H.~E., and Pammolli, F., Reputation and impact in
  academic careers, \emph{Proceedings of the National Academy of Sciences of
  the United States of America} \textbf{111} (2014) 15316--15321.

\bibitem{phelan}
Phelan, T.~J., Striking the mother lode in science: The importance of age,
  place, and time, \emph{The Journal of Higher Education} \textbf{65} (1994)
  627--629.

\bibitem{prozesky2008career}
Prozesky, H., A career-history analysis of gender differences in publication
  productivity among {S}outh {A}frican academics, \emph{Science \& Technology
  Studies} \textbf{21} (2008) 47--67.

\bibitem{Prpi2002}
Prpi{\'{c}}, K., Gender and productivity differentials in science,
  \emph{Scientometrics} \textbf{55} (2002) 27--58.

\bibitem{DBLP:series/lnsn/Reitz013}
Reitz, F. and Hoffmann, O., Learning from the past: An analysis of person name
  corrections in {DBLP} collection and social network properties of affected
  entities, in \emph{2010 International Conference on Advances in Social
  Networks Analysis and Mining} (2010), pp. 9--16.

\bibitem{DBLP:conf/ercimdl/ReutherWLWK06}
Reuther, P., Walter, B., Ley, M., Weber, A., and Klink, S., Managing the
  quality of person names in {DBLP}, in \emph{Research and Advanced Technology
  for Digital Libraries, 10th European Conference, {ECDL}} (2006), pp.
  508--511.

\bibitem{Rhoten200756}
Rhoten, D. and Pfirman, S., Women in interdisciplinary science: Exploring
  preferences and consequences, \emph{Research Policy} \textbf{36} (2007)
  56--75.

\bibitem{Sarigoel2014}
Sarig\"{o}l, E., Pfitzner, R., Scholtes, I., Garas, A., and Schweitzer, F.,
  Predicting scientific success based on coauthorship networks, \emph{EPJ Data
  Science}  (2014).

\bibitem{servia2015evolution}
Servia-Rodr{\'\i}guez, S., Noulas, A., Mascolo, C., Fern{\'a}ndez-Vilas, A.,
  and D{\'\i}az-Redondo, R.~P., The evolution of your success lies at the
  centre of your co-authorship network, \emph{PloS ONE} \textbf{10} (2015)
  e0114302.

\bibitem{shalizi_homophily_2011}
Shalizi, C.~R. and Thomas, A.~C., Homophily and contagion are generically
  confounded in observational social network studies, \emph{Sociological
  Methods \& Research} \textbf{40} (2011) 211--239.

\bibitem{Sinatraaaf5239}
Sinatra, R., Wang, D., Deville, P., Song, C., and Barab{\'a}si, A.-L.,
  Quantifying the evolution of individual scientific impact, \emph{Science}
  \textbf{354} (2016).

\bibitem{Spelke05}
Spelke, E.~S., Kinzler, K., and Shusterman, A., Sex differences in intrinsic
  aptitude for mathematics and science? {A} critical review, \emph{American
  Psychologist} \textbf{60} (2005) 950--958.

\bibitem{Stack2004}
Stack, S., Gender, children and research productivity, \emph{Research in Higher
  Education} \textbf{45} (2004) 891--920.

\bibitem{Tang:08KDD}
Tang, J., Zhang, J., Yao, L., Li, J., Zhang, L., and Su, Z., Arnetminer:
  Extraction and mining of academic social networks, in \emph{Proceedings of
  the 14th ACM SIGKDD International Conference on Knowledge Discovery and Data
  Mining}, KDD '08 (ACM, 2008), pp. 990--998.

\bibitem{Ugander17042012}
Ugander, J., Backstrom, L., Marlow, C., and Kleinberg, J., Structural diversity
  in social contagion, \emph{Proceedings of the National Academy of Sciences of
  the United States of America} \textbf{109} (2012) 5962--5966.

\bibitem{vanDerLee2015}
van~der Lee, R. and Ellemers, N., Gender contributes to personal research
  funding success in the netherlands, \emph{Proceedings of the National Academy
  of Sciences} \textbf{112} (2015) 12349--12353.

\bibitem{Vasileiadou20091260}
Vasileiadou, E. and Vliegenthart, R., Research productivity in the era of the
  internet revisited, \emph{Research Policy} \textbf{38} (2009) 1260--1268.

\bibitem{watts_collective_1998}
Watts, D.~J. and Strogatz, S.~H., Collective dynamics of ``small-world''
  networks, \emph{Nature} \textbf{393} (1998) 440--442.

\bibitem{Way2017}
Way, S.~F., Morgan, A.~C., Clauset, A., and Larremore, D.~B., The misleading
  narrative of the canonical faculty productivity trajectory, \emph{ArXiv
  e-prints} \textbf{1612.08228} (2016).

\bibitem{wenneras2001nepotism}
Wenneras, C. and Wold, A., Nepotism and sexism in peer-review, \emph{Nature}
  \textbf{387} (1997) 341--343.

\bibitem{West2013}
West, J.~D., Jacquet, J., King, M.~M., Correll, S.~J., and Bergstrom, C.~T.,
  The role of gender in scholarly authorship, \emph{PLoS ONE} \textbf{8} (2013)
  e66212.

\bibitem{West2006}
West, M.~S. and Curtis, J.~W., {AAUP} faculty gender equity indicators,
  Technical Report (2006).

\bibitem{westermann_gender_2005}
Westermann, O., Ashby, J., and Pretty, J., Gender and social capital: {The}
  importance of gender differences for the maturity and effectiveness of
  natural resource management groups, \emph{World Development} \textbf{33}
  (2005) 1783--1799.

\bibitem{white_identity_2008}
White, H.~C., \emph{Identity and Control: {H}ow Social Formations Emerge}
  (Princeton University Press, 2008).

\bibitem{wickware1997along}
Wickware, P., Along the leaky pipeline, \emph{Nature} \textbf{390} (1997)
  202--20.

\bibitem{Wuchty1036}
Wuchty, S., Jones, B.~F., and Uzzi, B., The increasing dominance of teams in
  production of knowledge, \emph{Science} \textbf{316} (2007) 1036--1039.

\bibitem{yang2012structure}
Yang, J. and Leskovec, J., Structure and overlaps of communities in networks,
  \emph{ArXiv e-prints} \textbf{1205.6228v2} (2012).

\bibitem{Xiao2016}
Zeng, X. H.~T., Duch, J., Sales-Pardo, M., Moreira, J.~A., Radicchi, F.,
  Ribeiro, H.~V., Woodruff, T.~K., and Amaral, L. A.~N., Differences in
  collaboration patterns across discipline, career stage, and gender,
  \emph{PLoS Biology} \textbf{14} (2016) e1002573.

\bibitem{SIM:SIM486}
Zheng, B., Summarizing the goodness of fit of generalized linear models for
  longitudinal data, \emph{Statistics in Medicine} \textbf{19} (2000)
  1265--1275.

\end{thebibliography}

\end{document}